\documentclass[12pt,a4paper]{article} 

\usepackage{epsfig}
\usepackage{amssymb}

\newcommand{\be}{\begin{equation}}
\newcommand{\en}{\end{equation}}
\newcommand{\bea}{\begin{eqnarray}}
\newcommand{\ena}{\end{eqnarray}}

\newcommand{\hbo}{\hbox to 1 true cm {\hfill } }
\newcommand{\tr}{\hbox{tr}}

\newcommand{\hs}{\hspace{0.9cm}}

\begin{document}

\vglue 1 truecm

\vbox{ UNITU-THEP-018/2001 
\hfill October 25, 2001
}
  
\vfil
\centerline{\large\bf Gluon propagators and quark confinement } 
  
\bigskip
\centerline{ K.~Langfeld$^a$, H.~Reinhardt$^b$ and J.~Gattnar  } 
\vspace{1 true cm} 

\centerline{ Institut f\"ur Theoretische Physik, Universit\"at 
   T\"ubingen }
\centerline{D--72076 T\"ubingen, Germany}
  
\vfil
\begin{abstract}

The gluon propagator is investigated in Landau and in maximal center gauge 
for the gauge group  SU(2) by means of lattice gauge simulations. 
We find that Gribov ambiguities arising from the implementation of 
Landau gauge have a small influence on the propagator. In agreement 
with previous findings, we obtain that the small momentum behavior is 
dominated by a mass, which is of order $(1.48 \pm 0.05) \, \sqrt{\sigma }$, 
where $\sigma $ is the string tension. By removing the confining vortices 
from the full Yang-Mills ensemble, we convert full YM-theory into a theory 
which does not confine quarks. We find that in the latter case 
the strength of the gluon propagator in the intermediate momentum range 
is strongly reduced. The spectral functions which reproduce the 
numerical data for the propagators are analyzed using a generalized 
Maximum Entropy Method.

\end{abstract}

\vfil
\hrule width 5truecm
\vskip .2truecm
\begin{quote} 

{\it keywords: gluon propagator, SU(2) lattice gauge theory, quark 
confinement, spectral function. }

$^a$ Supported by {\it Strukturfond 2000} of the University of T\"ubingen. 
$^b$ Supported by DFG Re856/4-1. 
\end{quote}
\eject

\section{ Introduction}

\vskip 0.3cm 
Nowadays, the general belief that QCD is the correct theory of strong 
interactions is supported by numerous high energy collisions 
experiments (see e.g.~\cite{ell96}). 
At high momentum transfers, at which the effective quark gluon coupling 
becomes small, predictions of perturbation theory nicely agree with the 
experimental data. On the other hand, this coupling becomes large 
and even diverges at low energies (Landau pole) thus prohibiting a 
perturbative treatment and pointing towards new phenomena. 
One might speculate that the strong increase of the effective coupling 
is the progenitor of quark confinement. Note, however, that a toy 
quark model which mimics the QCD perturbative behavior of the running 
coupling strength reveals that the new phenomenon at low energies is quark 
mass generation (which screens the Landau pole~\cite{la95})
rather than quark confinement. 

\vskip 0.3cm 
One prominent method to treat non-perturbative Yang-Mills theory is 
the numerical simulation of lattice gauge theory (LGT). 
LGT covers all non-perturbative effects and, in particular, bears 
witness of quark confinement (see e.g.~\cite{bali96}). 
Moreover, the numerical results suggest that, in certain gauges, 
topological defects act as quark confiners. The presence of these 
defects is independent of the coupling strength, and their properties 
are beyond the reach of perturbation theory. 
In the case of Abelian gauges~\cite{tho76}, these defects are 
chromomagnetic monopoles, which condense and give rise to the dual 
Meissner effect as confinement scenario~\cite{cea96,bali96,lam00} 
(for a caveat see~\cite{lam00}). A generalization 
of the dual Meissner effect to the non-abelian case was put forward 
in~\cite{la00a} to resolve the neutral particle problem. 
In the Center gauges~\cite{deb97}, the topological defects are center 
vortices. The mechanism of quark confinement can be understood as a 
percolation of center vortices which acquire physical 
relevance in the continuum limit~\cite{la98}. Lattice gauge calculations 
also provide an intuitive picture in terms of vortex physics for the 
deconfinement phase transition at finite temperatures~\cite{la99}. 
Reducing the full Yang-Mills 
configurations to their vortex content still yields the full string 
tension~\cite{deb97}. Vice versa, removing these vortices from the 
Yang-Mills ensemble results in a vanishing string tension~\cite{deb97}
and in addition to a restoration of chiral symmetry~\cite{for99}. 

\vskip 0.3cm 
Unfortunately, 
simulations of LGT including dynamical quarks are still cumbersome 
despite the recent successes by improved algorithms~\cite{kap92} and the 
increase of computational power. Moreover, at the present stage systems 
at finite baryon densities are hardly accessible in the realistic case of 
an SU(3) gauge group~\cite{bar98} (for recent successes see~\cite{eng99}). 
These deficiencies are overcome by a second non-perturbative method 
to treat Yang-Mills theory: the approach by the Dyson-Schwinger 
equations (DSE). By contrast to the lattice formulation, the DSE approach 
can easily deal with dynamical quarks and, furthermore, can be easily 
extended to finite baryon densities~\cite{rob00}. Moreover, it can be used 
to study hadron 
phenomenology~\cite{rob94,alk00}. The disadvantage of the DSE approach is 
that it requires a truncation of the infinite tower of equations, and that 
this approximation is difficult to control and to improve systematically. 
In addition, the DSE approach needs gauge fixing which is obscured by 
Gribov copies. Whether the standard Faddeev-Popov method of gauge fixing 
is appropriate in non-perturbative studies, is still under 
debate~\cite{bau98}. 

\vskip 0.3cm 
In view of the importance of the DSE approach for the understanding of 
hadron physics~\cite{rob94,alk00}, it is highly interesting to learn more 
about the impact of truncating the Dyson tower of equations. Since 
the gluon and ghost propagators are essential ingredients of the quark DSE, 
these propagators are of particular interest. 

\vskip 0.3cm 
In this paper, we will address the gluon propagator in Landau gauge for 
the simpler case of pure SU(2) lattice gauge theory. 
The lattice result will be compared with the one provided by the solution 
of the coupled ghost-gluon Dyson equation~\cite{sme97,atk98,atk98b,blo01}. 
By removing the confining center vortices 
from the SU(2) Yang-Mills ensemble, we will focus on 
the information on quark confinement which might be encoded in the gluon 
propagator. High accuracy numerical data for the latter 
are obtained by means of a new numerical method superior to 
existing techniques. 

\vskip 0.3cm 
The organization of the paper is as follows: In section 2, we present 
the lattice definition of the gluon propagator. For making contact with 
the ab initio continuum formulation of Yang-Mills theory, we will define 
the gluon field from the adjoint links. We furthermore outline in this 
section the gauge fixing procedure. In section 3, we present our 
numerical results for the gluon propagator in the Landau gauge and the 
maximal center gauge, respectively. In section 4, we discuss the 
gluonic spectral function which is extracted from the gluon propagator 
by using (an extension of) the maximal entropy method. Finally, our 
conclusions are presented in section 5.

\section{ The lattice approach to the gluon propagator } 

In this section, we will extract the gluon propagator of the continuum 
Yang-Mills theory by considering the continuum limit of 
the lattice gauge theory. Thereby, we will carefully examine the 
relation between the lattice link variables and the gauge potential.

\subsection{ The lattice definition of the gluon field } 

Before identifying the gluonic degrees of freedom in the lattice 
formulation, we briefly recall the definition of the gluon field 
in continuum Yang-Mills theory. For simplicity, we will consider the 
case of SU(2) Yang-Mills theory. 

\vskip 0.3cm 
Under a gauge transformation of the fundamental matter field, i.e. 
\be 
q(x) \rightarrow 
q^\prime (x) \; = \; \Omega \, (x) \; q(x) \; , \hbo 
\Omega (x) \in {\mathrm SU(2) } \; , 
\label{eq:1} 
\en 
the gluon fields transforms as 
\bea 
A^{a \, \prime }_{\mu }(x) &=& O^{ab}(x) \,  A^b _\mu (x) 
\; + \; \frac{1}{2} \epsilon^{aef} \, O^{ec}(x) \, \partial _\mu O^{fc} \; , 
\label{eq:3a} \\ 
O^{ab}(x) &:=& 2 \, \tr \bigl\{ \Omega (x) \, t^a \, 
\Omega ^\dagger (x) \, t^b \bigr\} \; , \hbo 
O^{ab}(x) \, \in \, {\mathrm SO(3) } \; . 
\label{eq:3} 
\ena 
Let us stress that the gluon fields transform according 
to the {\it adjoint} representation while the matter fields are 
defined in the fundamental representation. 

\vskip 0.3cm 
Let us compare these definitions of fields with the ones in LGT. 
In LGT, a discretization of space-time with a lattice spacing $a$ 
is instrumental. The 'actors' of the theory are SU(2) matrices 
$U_\mu (x)$ which are associated with the links of the lattice. 
These link variables transform under gauge transformations as 
\be 
U^\prime _\mu (x) \; = \; \Omega (x) \, U_\mu (x) \, \Omega ^\dagger  
(x+\mu)  \; \hbo \Omega (x) \in {\mathrm SU(2) } \; . 
\label{eq:4} 
\en 

For comparison with the ab initio continuum formulation, we also 
introduce the adjoint links 
\bea 
{\cal U } _{\mu }^{ab} (x) &:=&  2 \, \tr \bigl\{ U_\mu (x) \, t^a \, 
U^\dagger _\mu (x) \, t^b \bigr\} \; , \hbo 
t^a = \frac{1}{2} \tau ^a 
\label{eq:5} \\ 
{\cal U } _{ \mu }^{\prime }(x) &:=&  O(x) \, {\cal U } _{\mu } (x) \, 
O^T (x+\mu) \; , \hbo O(x) \in {\mathrm SO(3) } \; , 
\ena 
where $O(x)$ was defined in (\ref{eq:3}), and $\tau ^a$ are the 
Pauli matrices. 

\vskip 0.3cm 
In order to define the gluonic fields from lattice configurations, 
we exploit the behavior of the (continuum) gluon fields under 
gauge transformations (see~(\ref{eq:3a})), and identify the lattice 
gluon fields $A_\mu ^a (x)$ as algebra valued fields of the adjoint 
representation, i.e. 
\be 
{\cal U } _{\mu }^{cd} (x) \; =: \; \biggl[ \exp \bigl\{ \hat{t} ^f  
\, A^f_\mu (x) \, a \bigr\} \, \biggr] ^{cd} \; , 
\hbo \hat{t} ^f _{ac} := \epsilon ^{afc} \; , 
\label{eq:6} 
\en 
where the total anti-symmetric tensor $\epsilon ^{abc}$ 
is the generator of the SU(2) group in the adjoint representation, and 
where $a$ denotes the lattice spacing.

\vskip 0.3cm 
For later use, it is convenient to have an explicit formula for the 
(lattice) gluon fields $A_\mu ^a (x)$ defined by (\ref{eq:6}) 
in terms of the 
SU(2) link variables $U_\mu (x)$. Usually, these links are given  in terms 
of four--vectors of unit length 
\be 
U_\mu (x) \; = \; u_\mu ^0(x) \, + \, i \, \vec{u}_\mu (x) \, 
\vec{\tau} \; , \hbo \bigl[ u^0_\mu (x)\bigr]^2 \, + \, 
\bigl[ \vec{u}_\mu (x)\bigr]^2 \; = \; 1 \; . 
\label{eq:7} 
\en 
Inserting this representation for $U_\mu (x)$ into (\ref{eq:5}), 
we expand (\ref{eq:5}) and (\ref{eq:6}), respectively, in powers 
of the lattice spacing using 
$$ 
\vec{u}_\mu (x) = {\cal O }(a) \; , \hbo 
(u_\mu ^0)^2 (x) = 1 - {\cal O}(a^2) \; , 
$$ 
Comparing the order ${\cal O}(a)$, we find 
\be 
A^b_\mu (x) \, a \; + \; {\cal O}(a^2) \; = \; 2 \, u^0_\mu (x) 
\, u^b_\mu (x) \; , \hbox to 5cm{ without summation over  } \, \mu \; .
\label{eq:8} 
\en 
Let us emphasize that this gauge field $A^b_\mu (x) $ has been 
defined by the adjoint link (\ref{eq:6}). 
As a consequence,  the representation (\ref{eq:8}) is invariant under 
a non-trivial $Z_2$ center transformation, i.e. $ U_\mu (x) 
\rightarrow - \, U_\mu (x) $.  

\vskip 0.3cm 
Let us contrast our definition (\ref{eq:8}) with the previous 
definition of the gauge field from the fundamental link, i.e. 
\be 
U_\mu (x) \; = \; \exp \biggl\{ i \, a \bar{A}^b_\mu t^b \biggr\} \; , \hbo 
a \bar{A}^b_\mu  \; = \; 2 \, u^b_\mu (x) \; + \; {\cal O}(a^2) , 
\label{eq:old} 
\en 
where one must assume that the link field $U_\mu (x) $ is close to the 
unit element, i.e. $u_\mu ^0 (x) = 1 - {\cal O}(a^2)$. Indeed, 
the gluon field $\bar{A}^b _\mu (x)$ changes sign under a non-trivial 
center transformation. The previous definition of the gauge field 
$\bar{A}^b_\mu $ therefore contains information on center elements 
and coset fields as well. Here, we propose to disentangle the 
information carried by center elements and coset fields $A^b_\mu (x)$ 
and to study their correlations separately. In the following, we will 
present the correlation function of the coset ''gluon'' fields 
$A^b_\mu (x)$.

\subsection{ Gauge fixing }
\label{sec:gf} 

Calculation of the gluon propagator requires gauge fixing. In order 
to be able to compare with the Dyson-Schwinger approach, we firstly 
use the (lattice) Landau gauge condition 
\be 
\Omega (x): \; \sum _{\{x\}, \mu } \tr \, U^\prime _\mu (x) 
\rightarrow \, \mathrm{max} \; , 
\label{eq:9} 
\en 
where $U^\prime _\mu (x) $ is the gauged transformed link (\ref{eq:4}). 
Gauge transformation $\Omega (x)$ which are determined by (\ref{eq:9}) 
bring the SU(2) link elements as close as possible to the unit 
element. Decomposing the link variable in this gauge as 
\be 
U^\prime _\mu (x) \; = \; Z_\mu (x) \, \exp \biggl\{ i A^b_\mu (x) 
\, t^b \, a \biggr\} \; , 
\label{eq:10} 
\en 
where 
\be 
Z_\mu (x) \; = \; \mathrm{sign} \, \tr U_\mu(x) \; \in \, \{-1,+1\} \; , 
\hbo 
\sqrt{ \sum _b A^b_\mu  A^b_\mu } \in [0, \frac{ \pi }{a} ] \; , 
\label{eq:10a} 
\en 
one observes that in (lattice) Landau gauge (\ref{eq:9}) the role 
of the $Z_2$ center elements are de-emphasized (almost all $Z_\mu (x)$ 
are $1$) and most of the physics is contained in the adjoint field 
$A^b_\mu (x) $ (\ref{eq:8}). 
For this reason, we do not expect a vastly different gluon propagator 
when the more standard definition of the lattice gluon fields, i.e. 
$\bar{A}^b_\mu (x)$ (see (\ref{eq:old})), 
in terms of the fundamental links $U_\mu(x)$ is used~\cite{cuc97,bon01}. 

\vskip 0.3cm 
In the case of the gauge (\ref{eq:9}), the gluon fields (\ref{eq:8}) 
satisfy the familiar Landau gauge\footnote{ 
Strictly speaking, the eq.(\ref{eq:11}) holds up to singular points 
where $Z_\mu(x)=-1$ (see (\ref{eq:10})). These singularities are 
subject of future work~\cite{la01}.}
\be 
\partial _\mu A^{\prime \, a }_\mu (x) \; = 0 \; . 
\; , 
\label{eq:11} 
\en 
It is this gauge which is used in the Dyson-Schwinger studies. 
In this approach, the gauge condition (\ref{eq:11}) 
is implemented by means of the Faddeev-Popov method, thereby relying 
on the assumption that the Faddeev-Popov determinant represents the 
probabilistic weight of the gauge obit specified by its representative 
$A^{\prime \, a }_\mu (x) $. This method is correct 
if the gauge condition picks a unique solution $\Omega (x)$ of (\ref{eq:11}) 
for a given field $A^a_\mu (x)$. Unfortunately, the Landau gauge 
condition generically admits several solutions depending on the 
''background field'' $A_\mu^b(x)$ (Gribov ambiguity). Further restrictions 
on the space of possible solutions $\Omega (x)$ are required~\cite{zwa94}. 
It was argued in~\cite{bau98} that the Faddeev-Popov method is not always 
justified if Gribov ambiguities are present. 

\vskip 0.3cm 
Let us contrast the continuum gauge fixing with its lattice analog. 
In a first step, link configurations $U_\mu(x)$ are generated by means 
of the gauge invariant action without any bias to a gauge condition. 
In a second step, the gauge-fixed ensemble is obtained by 
adjusting the gauge matrices $\Omega (x)$ (see~(\ref{eq:4})) 
until the gauged link ensemble satisfies the gauge condition (\ref{eq:9}). 
When one representative of each gauge orbit is picked by this 
procedure, one must determine with which weight this representative 
contributes to the observable of interest. In the absence of Gribov 
ambiguities, this weight factor is given by the Faddeev Popov determinant
(see e.g.~\cite{weinb}). The important point is that in the lattice 
calculation the Faddeev Popov determinant needs not to be explicitly 
evaluated. This is because the unbiased generation of gauge field 
configurations which are subsequently transformed into the desired gauge 
produces each configuration of the gauge orbit with equal weight. Hence, 
this procedure automatically produces the proper probability distribution 
of the gauge fixed sub-manifold to which the representative belongs. 
Further details of the numerical approach to gauge fixing can be 
found in appendix \ref{app:a}. 

\vskip 0.3cm 
However, this does not dispense us from dealing with the Gribov
ambiguity. Let us illustrate this point: 
The naive Landau gauge condition for the gluon field (\ref{eq:11}) 
is satisfied if we seek an {\it extremum} (instead of the maximum) 
of the variational condition 
(\ref{eq:9}). If we restrict the variety of solutions $\Omega (x)$ 
which extremize (\ref{eq:9}) to those solutions which {\it maximize } 
the functional (\ref{eq:9}), we confine ourselves to the case where 
the Faddeev-Popov matrix is positive semi-definite. The 
corresponding fraction of the configuration space of gauge fixed 
fields $A^{\prime \, b} _\mu $ 
is said to lie within the first Gribov horizon. 
However, there is still a variety of possible gauge transformations 
$\Omega (x)$ which all correspond to local maxima of the functional 
(\ref{eq:9}). A conceptual solution which resolves this residual 
Gribov ambiguity 
is to restrict the configuration space of gauge fixed fields 
$A^{\prime \, b} _\mu $ to the so-called {\it fundamental modular 
region}. In the lattice simulation, this amounts to picking the 
{\it global maximum } of the variational condition (\ref{eq:9}). In 
practice, finding the global maximum is a highly non-trivial task.
A numerical algorithm 
which obtains the gauge transformation matrices $\Omega (x)$ from 
the condition (\ref{eq:9}) might fail to locate the global maximum, and 
the numerical simulation might still sample a particular set of local 
maxima. Different algorithms might lead to different 
local maxima, and, hence, implement different gauges. 
A comprehensive study of the Gribov problem using lattice methods can be 
found in~\cite{giu01}. 

\vskip 0.3cm 
Here, we will study two extreme 
cases of gauge fixing: firstly, we will implement a gauge by means of 
an iteration over-relaxation algorithm which almost randomly averages over 
the {\it local maxima} of the variational condition (\ref{eq:9}) (IO gauge). 
This is the standard gauge fixing algorithm commonly used by the 
community (for technical details, see e.g.~\cite{giu01}). 
The resulting gluon propagator will then be compared with the gluon propagator 
of a gauge where a simulated annealing algorithm searches for the {\it global 
maximum} (SA gauge). In the latter case, we used the algorithm outlined 
in~\cite{sum96}. 
We stress that it is not granted that the SA algorithm finds the global 
maximum, i.e. that the gauge is fixed to the fundamental modular region. 
Instead, another set of local maxima is randomly chosen. The SA approach 
generically yields higher values of the gauge fixing functional than 
the IO algorithm. 
It is the present state of the art to study the degree with which the 
observable of interest depends on the choice of the these two extreme cases
of gauge fixing. This will be done below. 
It will turn out that the gluon propagators of both 
gauges agree within statistical error bars. These observations indicate 
that the gluon propagator does not depend on the subset of 
configurations which we choose from the first Gribov domain.

\subsection{ Form factor calculations } 

The link configurations are generated using the Wilson action. 
We refrain from using a ''perfect action'' since we are interested in 
the gluon propagator in the full momentum range; simulations 
using perfect actions recover a good deal of continuum physics 
at finite values of the lattice spacing at the cost of a non-local action. 
For practical simulations, perfect actions are truncated which becomes 
an un-justified approximation at high energies where the full non-locality of 
the action must come into play. 

\vskip 0.3cm
In the present paper, most of the calculations were performed using a 
$16^3 \times 32$ lattice. The dependence 
of the lattice spacing on $\beta $ (renormalization), i.e. 
\be 
\sigma a^2 (\beta ) \; = \; 0.12 \, \exp \biggl\{ - \frac{ 6 \pi^2}{11} 
\, \bigl( \beta - 2.3 \bigr) \biggl \} \, , \hbo 
\sigma := (440 \,{\mathrm MeV}) ^2 \; , 
\label{eq:20} 
\en 
is appropriate for $\beta \in [2.1, 2.6]$ for the achieved numerical
accuracy. Given (\ref{eq:20}), it is straightforward to calculate the 
extension of the lattice in one direction, i.e $L_x= N_x a(\beta )$, 
where $N_x$ is the number of lattice points in $x$-direction. 
In order to estimate the momentum range which is covered by the actual 
simulation, it is convenient to introduce the UV-cutoff by 
$\Lambda = \pi / a(\beta )$. Table~\ref{tab:1} provides $L_t$ and 
$\Lambda $ for $N_t=32$. 

\vskip 0.3cm
Once gauge-fixed ensembles are obtained by implementing a variational 
gauge condition (see discussion of previous section), the gluon 
propagator is calculated using 
\be 
D^{ab}_{\mu \nu } (x-y) \; = \; \bigl\langle A^a_\mu (x) \, 
A^b_\nu (y) \, \bigr\rangle _{MC} \; , 
\label{eq:21} 
\en 
where $A^a_\mu (x)$ is defined in terms of the coset part of the 
link (see (\ref{eq:8})). 
The Monte-Carlo average is taken over 200 properly thermalized 
gauge configurations. Of particular interest is the Fourier transform 
of this propagator which is defined by 
\be 
D_{\mu \nu }^{ab} (\hat{p}) \; = \; a^4 \sum_x  D^{ab}_{\mu \nu } (x) 
\; \exp \bigl\{ i \hat{p}x \bigr\} \; , \hbo 
\hat{p}_k := \frac{ 2 \pi }{ N_k a } n_k \; , 
\label{eq:22a} 
\en 
where $n_k$ labels the Matsubara mode in $k$-direction and where 
$N_k$ is the number of lattice points in this direction. It is also 
convenient to define the lattice momentum $p$ by 
\be 
p_k \; := \; \frac{2}{a(\beta )} 
\, \sin \left( \frac{ \pi }{N_k} \, n_k \right) \; , 
\label{eq:p1} 
\en 
which coincides with the Matsubara momentum $\hat{p}_k$ in (\ref{eq:22a}) 
in the limit $n_k \ll N_k $. The definition (\ref{eq:p1}) has the 
advantage that e.g.~the free lattice propagator takes the familiar 
form $1/p_k p_k$. 

\vskip 0.3cm 
From perturbative Yang-Mills theory, one expects that the deviation 
of the full propagator from the free one is logarithmically small 
for large $p^2$. 
In order to work out the non-trivial information of the Yang-Mills 
dynamics on the propagator, it is useful to introduce the gluon form 
factor $F(p^2)$ by 
\be 
D(\hat{p}) \; =: \; \frac{F(p^2)}{p^2} \; , \hbo 
D(\hat{p}) \; := \; \sum _{a, \mu } D^{aa}_{\mu \mu} (\hat{p})
\label{eq:22} 
\en 
which measures the deviation of the full propagator from 
the free one. Since in Landau gauge the propagator is diagonal in color 
and transversal 
in Lorentz space, the form factor $F(p^2)$ contains the full information. 
\begin{table}
\caption{ Simulation parameters } 
\label{tab:1}
\begin{center}
\begin{tabular}{cccccc} \hline\hline
$\beta $ \hs & \hs 2.1 & \hs 2.2 & \hs 2.3 & \hs 2.4 & \hs  2.5  \\ 
L [fm] \hs & \hs 8.6 & \hs 6.6 & \hs 5.0 & \hs 3.8 & \hs 2.9 \\ 
$\Lambda $ [GeV] \hs & \hs 2.3 & \hs 3.0 & \hs 4.0 & \hs 5.2 & \hs 6.8 
\\ \hline\hline
\end{tabular}
\end{center}

\end{table}

\subsection{ Numerical method } 
In principle, the gluonic form factor can be obtained by the 
Fourier transform of the measured quantity (\ref{eq:21}). Note, however, 
that the propagator (\ref{eq:21}) is a rapidly decreasing function 
in coordinate space implying that the physical information at large distance 
$\vert x - y \vert $ is washed out by statistical noise and that, therefore, 
the information on the low momentum behavior is lost. To overcome this 
deficiency, it was proposed in~\cite{zwa91} to directly address the 
propagator in momentum space, i.e. 
\be 
D(\hat{p}) \; = \; \sum_{a, \mu } \frac{1}{N^2_\mathrm{all}} \Bigl\langle \; 
\biggl[ \sum _{x} A^a_\mu (x) \, \cos (\hat{p} x) \, \biggr] ^2 \; 
\; + \; \biggl[ \sum _{y} A^a_\mu (y) \, 
\sin (\hat{p} y) \, \biggr] ^2 \; \Bigr\rangle _{MC} \; , 
\label{eq:p2} 
\en 
where $N_\mathrm{all}$ is the number of lattice points. By the help 
of the translational invariance, i.e. 
$$
\bigl\langle A^a_\mu (x) \, A^b_\nu (y) \, \bigr\rangle _{MC} 
\; \propto \; f \bigl( x-y \bigr) \; , 
$$
one indeed finds from (\ref{eq:p2}) that 
\be 
D(\hat{p}) \; = \; \sum _{a, \mu } 
\sum _x \, \Bigl\langle \; A^a_\mu (x) \, A^a_\mu (0)  \; 
\Bigr\rangle _{MC} \; \cos \, \biggl( \, \hat{p} x \, \biggr)  \; . 
\label{eq:p3} 
\en 
Since the function $D(\hat{p})$ also contains the trivial factor 
$1/p^2$, which is also present in a free theory, a further increase of 
the numerical accuracy is achieved by directly addressing the 
form factor $F(p^2)$. For these purposes, we firstly choose the 
momentum transfer to $\hat{p} = (0,0,0,\hat{p}_4)^T$ without any loss 
of generality and define 
\be 
\Delta _t A_\mu (x) \; := \; A_\mu (x+a \, e_4) \, - \, A_\mu (x) \; ,  
\label{eq:p4} 
\en 
where $e_4$ is the unit vector in time direction. A straightforward 
calculation yields 
\be 
F(p_4^2) \; = \; \sum _{a, \mu } 
\frac{1}{N^2_\mathrm{all}} \Bigl\langle \; 
\biggl[ \sum _{x} \Delta _t A^a_\mu (x) \, \cos (\hat{p} x) \, \biggr] ^2 \; 
+ \biggl[ \sum _{y} \Delta _t A^a_\mu (y) \, 
\sin (\hat{p} y) \, \biggr] ^2 \; \Bigr\rangle _{MC}  , 
\label{eq:p5} 
\en 
where $p_4$ is the lattice momentum (\ref{eq:p1}) in time direction. 
By sake of the $\Delta _t$ operators in (\ref{eq:p5}) the free part, i.e. 
$1/p^2$, of the propagator $D(\hat{p})$ is precisely canceled, and 
we are left with the quantity of interest $F(p^2)$. 
It turns out that Monte-Carlo average (\ref{eq:p5}) allows for a
high precision measurement of the form factor.

\section{Numerical Results } 

\subsection{ Landau gauge } 

In this section, we will mainly use the iteration over-relaxation 
algorithm to implement the Landau gauge condition (\ref{eq:9}), and 
will refer to the corresponding gauge as IO-gauge. Later, we will also 
compare this gluonic form factor with one obtained in the SA-gauge 
where the gauge condition (\ref{eq:9}) is implemented by a simulated 
annealing algorithm. 

\vskip 0.3cm 
\begin{figure}[t]
\centerline{
\epsfxsize=0.5\linewidth 
\epsfbox{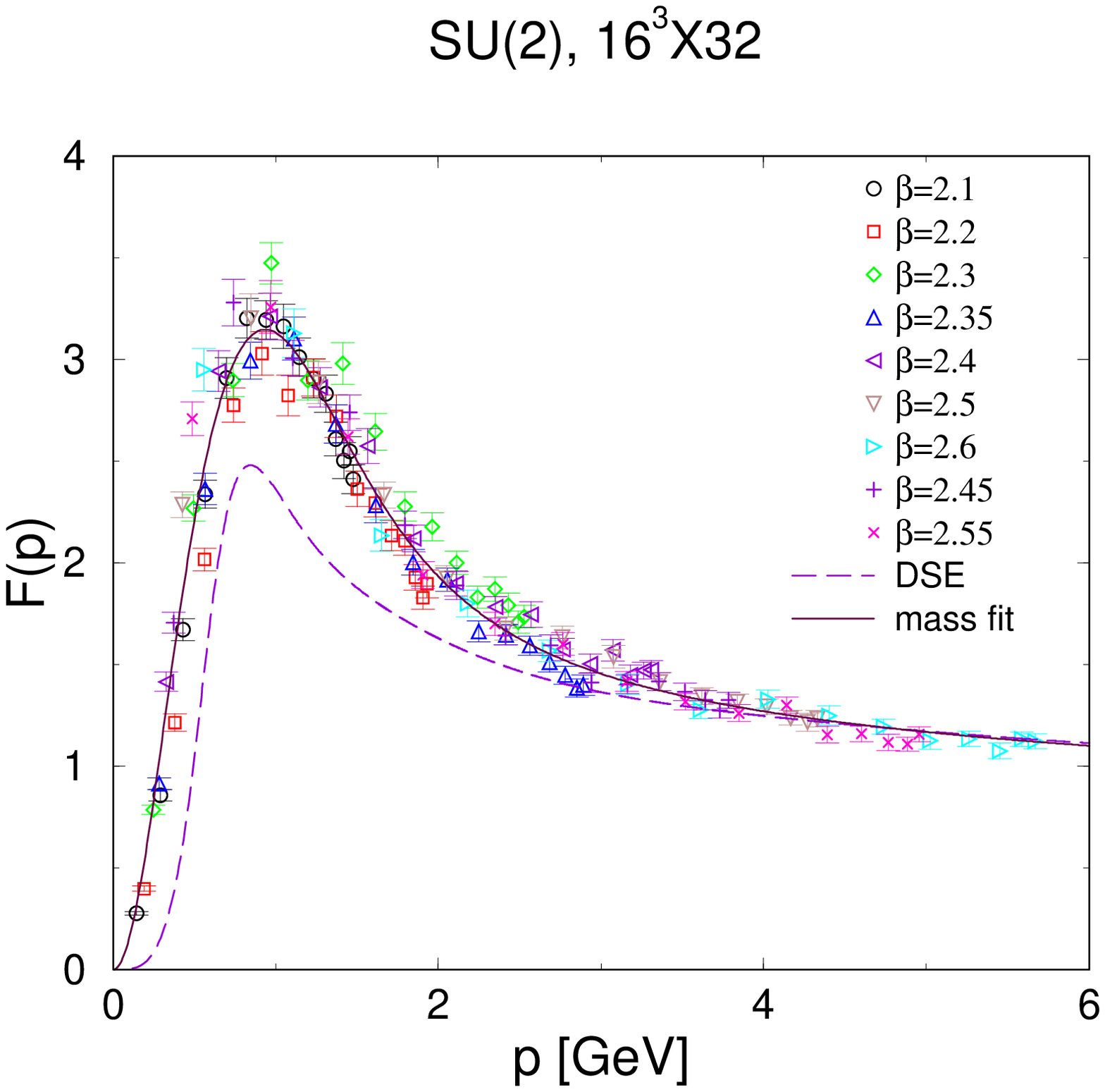}
\epsfxsize=0.55\linewidth 
\epsfbox{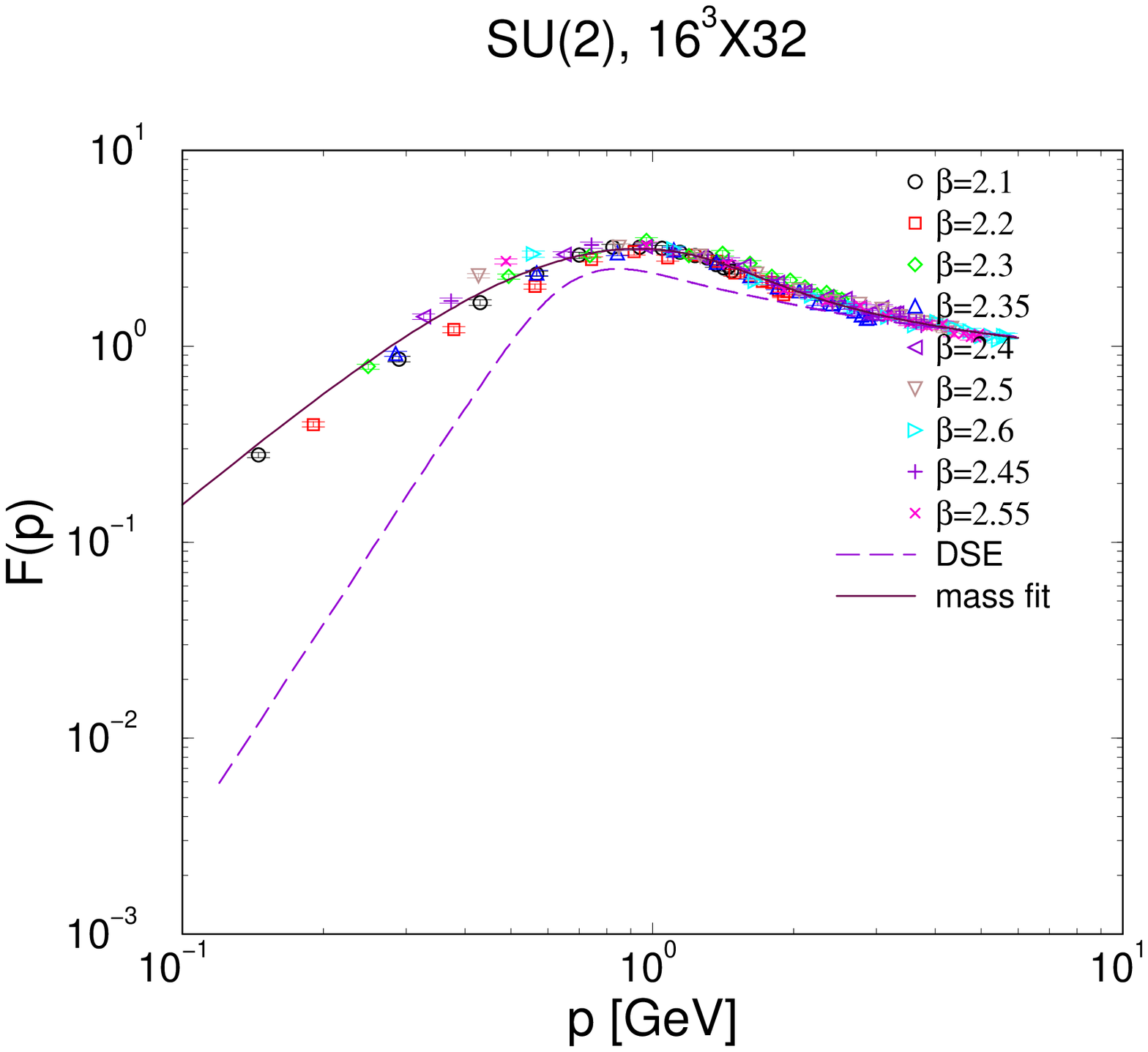}
}
\caption{The gluonic form factor $F(p^2)$ as function of the momentum 
   transfer (left panel: linear scale; right panel: log-log scale). 
   Also shown is the solution of the set of DSEs proposed 
   in~\cite{sme97} which have been solved for the case of 
   $SU(2)$~\cite{cfi01}.
}
\label{fig:1}
\end{figure}
In order to obtain the form factor $F(p^2)$ as function of the 
(lattice) momentum $p$ (\ref{eq:p1}), 200 properly thermalized configurations 
of a $16^3 \times 32$ lattice were used. Physical units for the momentum 
$p$ (\ref{eq:p1}) can be obtained by using (\ref{eq:20}). Calculations 
with different $\beta $-values correspond to simulations with a  
different UV-cutoff $\Lambda := \pi / a(\beta )$. 
In the first place, we obtain the un-renormalized form factor 
$F_B(p^2)$ as function of the momentum in physical units. The desired 
{\tt renormalized } form factor $F_R(p^2)$ is obtained via multiplicative 
renormalization, i.e. $F_R(p^2) = Z_3^{-1} F_B(p^2)$. Thereby, the gluonic 
wave function renormalization $Z_3(\Lambda )$ is chosen to yield a finite 
(given) value for the renormalized form factor $F_R$ at a fixed momentum 
transfer (renormalization point). In the following, we will suppress 
the subscripts of the form factors, and $F(p^2)$ always refers to the 
renormalized form factor. 

\vskip 0.3cm 
Figure~\ref{fig:1} shows the result for the gluonic form factor 
in the IO-Landau gauge. Our reference scale is the string tension, which 
we set to $\sigma = (440 \, \mathrm{MeV})^2$ in order to assign
physical units to momenta. 
The data for the renormalized form factor $F(p^2)$ which were obtained 
with different $\beta $-values nicely agree within numerical accuracy, 
thus signaling independence from the UV-cutoff and the lattice volumes 
in the range shown in table~\ref{tab:1}. 

\vskip 0.3cm 
At high momentum the lattice data are consistent with the behavior 
known from perturbation theory, 
\be 
F(p^2) \; \propto \; 1/ \left[ \mathrm{log} 
\frac{p^2}{\mu ^2 } \right]^{13/22} \; , 
\hbo p^2 \approx \mu ^2 \gg (1 \, \mathrm{GeV})^2 \; . 
\label{eq:24} 
\en 
Also shown in figure~\ref{fig:1} is the coarse grained 
''mass fit'' (${\cal N}$, 
$m_1$, $m_2$, $m_L$, $s$ fitted parameters; momentum $p$ and all 
mass scales in units of 1~GeV) 
\be 
F(p^2) \; = \; {\cal N} 
\frac{ p^2}{p^2+m_1^2} \biggl[ \frac{ 1 }{p^4+m_2^4} 
\; + \; \frac{s}{\left[ log \left(m_L^2 + p^2 
\right) \right]^{13/22} } \biggr] 
\label{eq:26} 
\en 
which nicely reproduces the lattice data within the statistical error bars 
for 
\bea 
{\cal N} &=& 8.1133 \; , \hbo 
m_1 = 0.64 \, \mathrm{GeV} \; , \hbo 
m_2 = 1.31 \, \mathrm{GeV} \; , \hbo 
\label{eq:26b} \\ 
s &=& 0.32                   \; , \hbo 
m_L = 1.23 \, \mathrm{GeV} \; . 
\nonumber 
\ena 
Note that the multiplicative renormalization only affects the 
normalization ${\cal N}$ implying that the other fitting parameters 
are renormalization group invariant quantities. 

\vskip 0.3cm
Let us point out that an effective gluonic mass was introduced 
in~\cite{con94} on phenomenological grounds. 
Gluonic masses were realized as electric and magnetic screening 
masses in the high temperature phase of SU(2) YM-theory~\cite{cuc01}, and 
were also reported in the maximal Abelian gauge~\cite{ame99} and 
in the  Laplacian Landau gauge~\cite{alex00}. 
Masses have been also reported for the case of a SU(3) gauge group 
in Landau gauge~\cite{bon01} and in Laplacian Landau 
gauge~\cite{alex00,alex01}. Note, 
however, that the description of the propagator in terms of two conjugate 
mass poles seems adequate for the Coulomb gauge~\cite{cuc00}.

\subsection{ Comparison with DSE solutions } 

Despite more than twenty years of successful hadron phenomenology 
originating from the quark~\cite{rob94,alk00} and 
gluon~\cite{man79,atk82,bro89} DSE, respectively, 
the coupled set of continuum 
DSEs for the renormalized gluon and ghost propagators has only 
recently been addressed in~\cite{sme97} and subsequently 
in~\cite{atk98,atk98b,blo01}. In ref.~\cite{sme97}, 
it was firstly pointed out that, 
at least for a specific truncation scheme, the gluon and ghost from factors 
satisfy scaling laws in the infra-red momentum range, 
in particular 
\be 
F(p^2) \; \propto \left[ p^2 \right]^{2 \kappa } \; , \hbo 
p^2 \ll \Lambda _{\mathrm QCD} ^2 \; . 
\label{eq:25}  
\en 
Depending on the truncation of the tower of Dyson equations and on 
the angular approximation of the momentum loop integral, one finds 
$\kappa = 0.92$~\cite{sme97} or $\kappa = 0.77$~\cite{atk98} 
or $\kappa \rightarrow 1$~\cite{atk98b}. The lattice data are consistent 
with $\kappa = 0.5$ corresponding to an infra-red screening by a gluonic 
mass (see figure \ref{fig:1} right panel). Interestingly, the 
prediction that the running coupling strength 
\be 
\alpha (p^2) \; = \; \alpha (\mu^2) \, F(p^2) \, G^2(p^2) \; , \hbo 
G(p^2): \, \hbox{ Ghost form factor } 
\label{eq:25b} 
\en 
approaches a constant in the limit $p^2 \rightarrow 0$ is independent 
of the truncation and approximations used 
in~\cite{sme97,atk98,atk98b,wat01}. 

\vskip 0.3cm 
A feature of full Yang-Mills theory is multiplicative renormalizability  (MR)
which guarantees that the form factors $F(p^2)$ and $G(p^2)$ 
might be rescaled independently to satisfy the renormalization condition
$F(\mu ^2)=1$, $G(\mu ^2)=1$ ($\mu $ renormalization point). 
Above, we have made use of this feature 
to obtain the renormalized form factor $F(p^2)$ from that lattice data 
for several values of $\beta $. It turned out that the truncations studied 
so far in~\cite{sme97,atk98,atk98b} violate multiplicative renormalizability 
to a certain extent (see~\cite{atk98} and \cite{blo01} for detailed 
discussions). Progress was made in~\cite{blo01} where a system of 
renormalized coupled gluon ghost DSEs were derived which manifestly 
reflect MR. Unfortunately, a self-consistent 
solution to this set of DSEs has not yet been obtained. 

\vskip 0.3cm 
For a quantitative comparison of our lattice data with the SU(2) 
DSE solution, we refer to the truncation scheme of~\cite{sme97} 
which incorporates Taylor-Slavnov identities to some extent. 
In this approach, the running coupling at the renormalization point 
$\mu $ serves as an input to assign physical units to the momenta. Assuming 
an approximate MR, the solution for $F(p^2)$ is then rescaled to 
satisfy $F(\mu^2)=1$. The final result is compared with the form factor 
obtained by our lattice calculations. 

\vskip 0.3cm 
The DSE approach requires a knowledge 
of the scale $\Lambda _{QCD}$ which is (to one loop accuracy) defined 
by (see e.g.~\cite{yndu}) 
\be 
F(p^2) \; = \; F(\mu ^2) \, \left( \frac{ \ln \left( \mu ^2 
/ \Lambda ^2_{QCD} \right) }{\ln \left( p ^2 
/ \Lambda ^2_{QCD} \right) } \right) ^{13/22} \; , 
\label{eq:27} 
\en 
which holds for $p^2 \approx \mu^2 \gg \Lambda ^2_{QCD}$. Defining 
\be 
\rho \; = \; \frac{\mu ^2}{ F(\mu ^2) } \frac{ d  F(p^2) }{ dp^2 } 
\biggr\vert _{p^2=\mu^2} 
\label{eq:28} 
\en 
and using (\ref{eq:27}), a straightforward calculation gives 
\be 
\Lambda ^2 _{QCD} \; = \; \mu ^2 \; \exp \left\{ \frac{13}{22 \, \rho } 
\right\} \; . 
\label{eq:29} 
\en 
For the parameter set (\ref{eq:26b}) which fits the lattice 
gluonic form factor with the string tension $\sigma = ( 440 \, 
\mathrm{MeV} )^2$ as reference scale, $\rho $ can be estimated from the 
fit function (\ref{eq:26}). The renormalization point, $\mu = 6 \, $GeV, 
was chosen to be part of the asymptotic momentum regime where 
both non-perturbative methods, i.e. LGT and the DSE approach, 
reproduce the known perturbative behavior. We finally obtain 
$\Lambda  _{QCD} \approx 889 \, \pm 10 \, $MeV, which is of the same order 
of magnitude than its SU(3) analog when the SU(3) gauge theory is equipped 
with the same string tension. At the renormalization point
$\mu = 6 \, $GeV, the running coupling constant for the SU(2) gauge theory 
is obtained by 
\be 
\alpha (\mu = 6 \, \mathrm{GeV}) \; = \; 
\frac{ 4\pi }{ 22/3 \; \ln (\mu^2 / \Lambda ^2_{QCD} ) } \; \approx \; 
0.449
\label{eq:30} 
\en 

\vskip 0.3cm 
Using (\ref{eq:30}) as input, the solution of the DSEs with the 
truncations of~\cite{sme97} was solved~\cite{cfi01} for the case of the 
SU(2) gauge group~\footnote{ We thank C.~Fischer for 
communicating his DSE solution for the SU(2) case prior to publication.} 
which yields the DSE result shown in figure~\ref{fig:1}. 
We find a qualitative agreement of the DSE solution with our 
lattice result. Given the varieties of exponents $\kappa $ (\ref{eq:25}) 
depending on the truncations of the DSEs, we do not expect 
a detailed agreement of the form factor close to zero momentum. 
We point out that the peak of the form factor at the intermediate 
momentum range is also observed in the DSE approach, although 
there are quantitative deviations concerning its steepness.

\subsection{ Gluon propagator and confinement } 

\begin{figure}[t]
\centerline{
\epsfxsize=0.53\linewidth 
\epsfbox{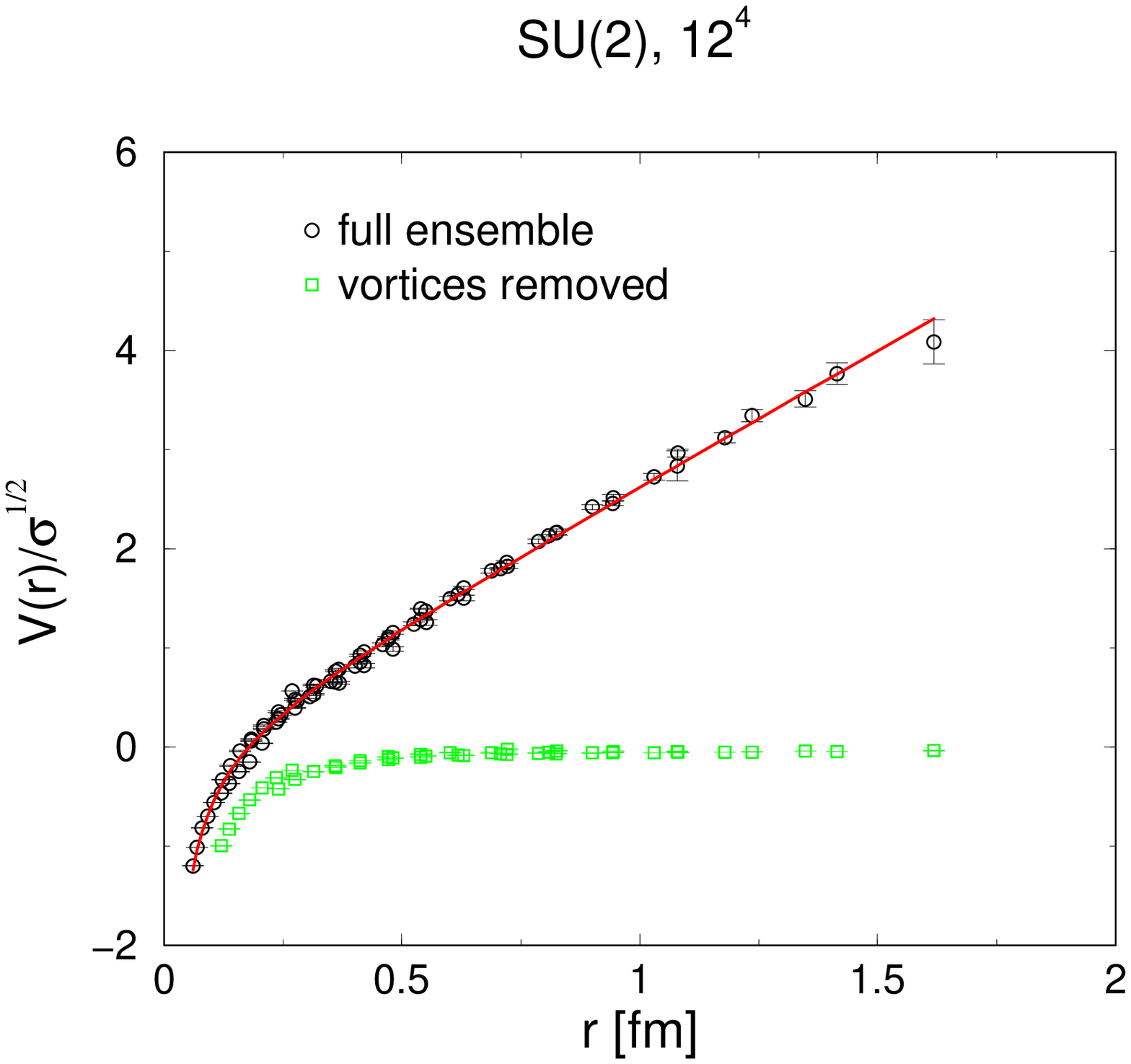}
\epsfxsize=0.5\linewidth 
\epsfbox{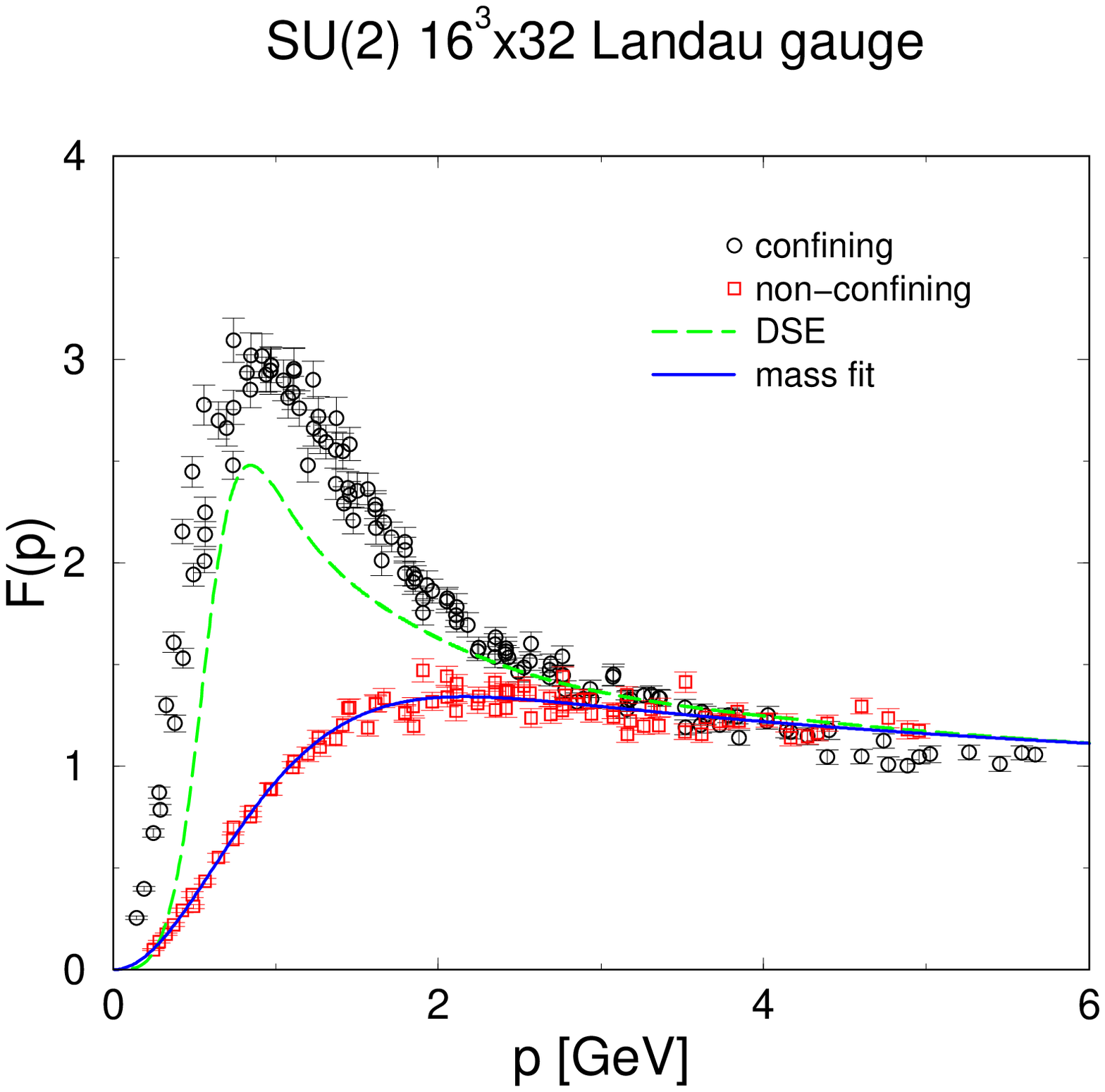}
}
\caption{ The static quark anti-quark potential (left panel) and 
   the corresponding gluonic form factors (right panel); DSE solution 
   from~\cite{cfi01}. } 
\label{fig:2}
\end{figure}
In the so-called maximum center gauges~\cite{deb97}, the role of the 
coset part of the links referred to as ''gluons'' 
for the infra-red physics is de-emphasized, and center vortices 
appear as physical degrees of freedom in the continuum limit~\cite{la98}. 
In this gauge these vortices act as the confiners of the theory: 
reducing the full Yang-Mills configurations to their vortex content still 
yields the full string tension~\cite{deb97}, and, vice versa, removing 
these vortices from the Yang-Mills ensemble results in a theory with 
a vanishing string tension~\cite{deb97,for99}. 

\vskip 0.3cm 
In order to get a handle on the information of quark confinement 
encoded in the gluon propagator in Landau gauge, we firstly modify 
the SU(2) Yang-Mills theory to a theory which does not confine quarks, 
and, secondly, we compare the form factor of the gluon 
propagator of the modified 
theory with the full SU(2) result (see figure~\ref{fig:1}). 

\vskip 0.3cm 
Let us briefly outline the numerical procedure for removing the 
confining center vortices from the Yang-Mills ensemble: firstly, 
we implement the maximal center gauge (MCG) condition  
\be 
\Omega (x): \; \sum _{\{x\}, \mu } \biggl[ \tr \, U^\prime _\mu (x) 
\biggr]^2 \rightarrow \, \mathrm{max} \; .
\label{eq:p10} 
\en 
by an iteration over-relaxation algorithm~\cite{deb97}, and obtain the 
$Z(2)$ vortex links by center projection
$$ 
U_\mu (x) \; \stackrel{MCG}{\rightarrow } U^{MCG}_\mu (x) 
\; \stackrel{proj}{\rightarrow } \; Z_\mu (x) \; = \; \mathrm{sign} 
\, \tr U^{MCG}_\mu (x) \; . 
$$
The elements $Z_\mu(x)$ span an $Z(2)$ gauge theory which contains 
the confining vortices as physical degrees of freedom. 
In order to remove the vortices from the full ensemble, we 
define configurations 
$$ 
U_\mu ^{mod} (x) \; := \; Z_\mu (x) \, U_\mu  (x) \; . 
$$ 
The configurations $ U_\mu ^{mod} (x) $ do not yield quark confinement 
any more: the static quark anti-quark potential calculated from 
ensembles $ U_\mu ^{mod} (x) $ (see figure \ref{fig:2} left panel) 
illustrates that a removal of the center vortices produces a 
non-confining theory.  

\vskip 0.3cm 
Finally, we implement 
the IO-Landau gauge condition on the configurations  $ U_\mu ^{mod} (x) $ 
and calculate the corresponding renormalized form factor by the 
procedure outlined in the previous section. Figure \ref{fig:2} right panel  
shows this gluonic form factor obtained from the modified ensemble. 
The striking feature is that the strength of the form factor in the 
intermediate momentum range is drastically reduced. 

\subsection{ Maximal center gauge } 

\begin{figure}[t]
\centerline{
\epsfxsize=0.5\linewidth 
\epsfbox{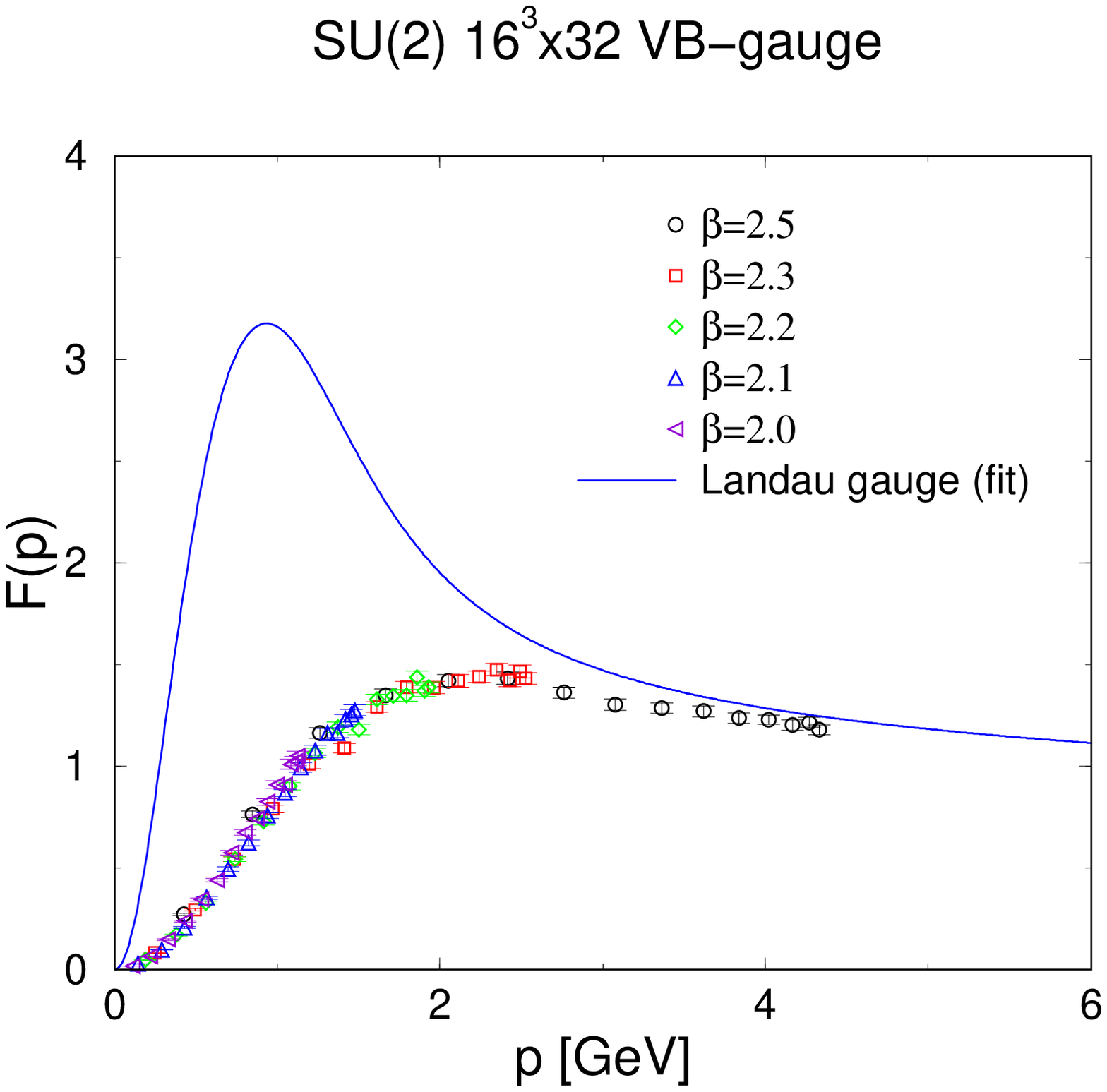}
\epsfxsize=0.52\linewidth 
\epsfbox{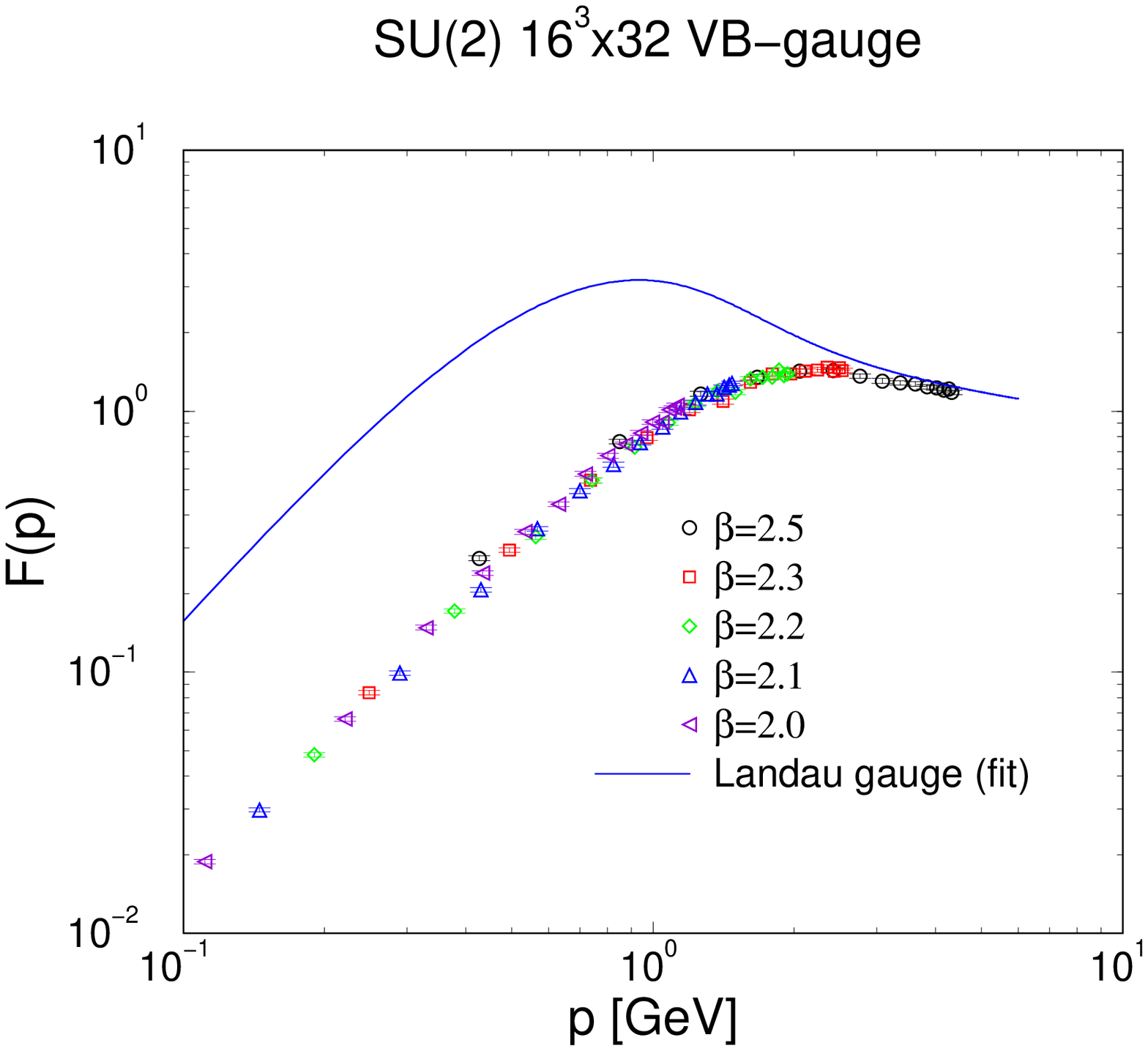}
}
\caption{The gluonic form factor $F(p^2)$ in MCG as function of the momentum 
   transfer (left panel: linear scale; right panel: log-log scale) 
   for the case of IO-MCG. 
}
\label{fig:3}
\end{figure}
In the maximal center gauge~\cite{deb97}-~\cite{for99}, important parts 
of the infra-red physics is shifted to the vortex degrees of freedom. It is 
therefore instructive to investigate the 
residual information carried by the gluonic form factor in this gauge 
(\ref{eq:p10}). 
We used an standard iteration over-relaxation algorithm~\cite{deb97} 
to determine the gauge matrices $\Omega (x)$ for a given ''background'' 
configuration $U_\mu (x)$. 

\vskip 0.3cm 
Once the gauge condition (\ref{eq:p10}) is implemented, we 
used the adjoint link to define the gluon field (see (\ref{eq:8})). 
The resulting gluon field satisfies the Landau gauge condition 
$\partial _\mu A_\mu ^b (x) =0$, since the MCG is equivalent to 
the adjoint Landau gauge 
\be 
\Omega (x): \; \sum _{\{x\}, \mu } \biggl[ \tr \, {\cal U } ^\prime 
_\mu (x) \biggr]^2 \rightarrow \, \mathrm{max} \; , 
\label{eq:adj} 
\en 
where ${\cal U } ^\prime _\mu (x)$ is the adjoint link (\ref{eq:5}). 

A thorough study of the MCG in the continuum limit was performed 
in~\cite{eng00}. One finds that in the 
continuum limit the gauge condition (\ref{eq:p10}) corresponds to  
a back ground gauge 
\be 
[\partial _\mu \, + \, i A^B_\mu (x), A_\mu (x) ] \; = \; 0 \; , 
\label{eq:p11} 
\en 
where the back ground gauge field $A^B_\mu (x)$ is an optimally chosen 
center vortex field (see~\cite{eng00} for details).  
In the absence of center vortices in the considered gauge field 
$A_\mu (x)$, the gauge (\ref{eq:p11}) coincides with the Landau gauge. 

\vskip 0.3cm 
The gluon form factor in MCG is shown in figure \ref{fig:3}. 
By comparing figures \ref{fig:1} and \ref{fig:3} we observe that 
the form factors in Landau gauge and MCG drastically differ in the 
intermediate region. This is because in Landau gauge most of the 
information is accumulated in the adjoint (coset) part of the links 
while in the MCG the part of the non-perturbative content is shifted 
to its center part. Furthermore, comparing figures \ref{fig:2} and 
\ref{fig:3} we find that the gluon propagator in MCG basically 
agrees with the one where the center vortices are removed and the 
Landau gauge is subsequently implemented. This underlines the observation 
that in MCG confinement physics is relegated to center vortices.

\subsection{ The Gribov noise } 

\begin{figure}[t]
\centerline{
\epsfxsize=0.5\linewidth 
\epsfbox{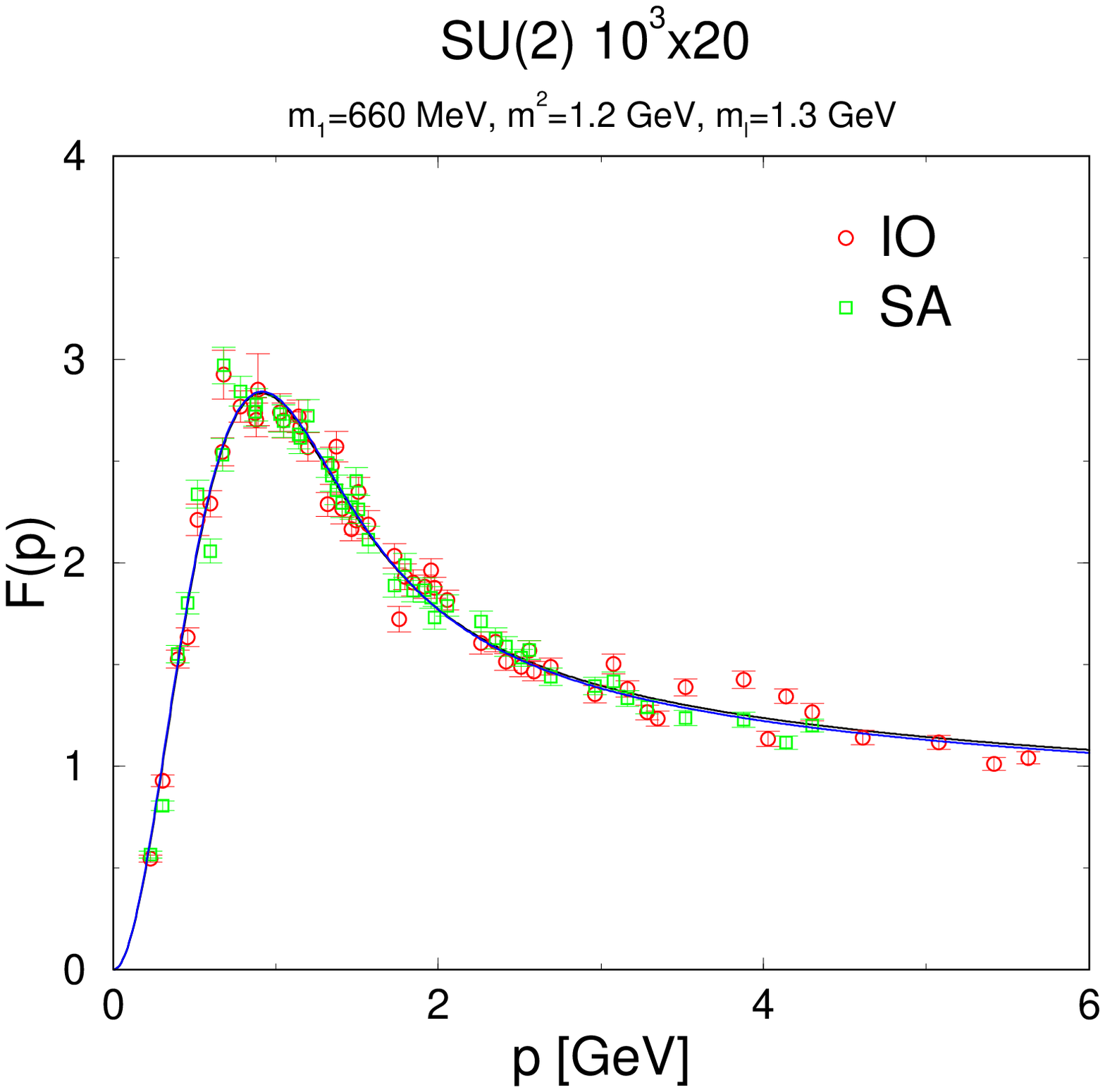}
\epsfxsize=0.5\linewidth 
\epsfbox{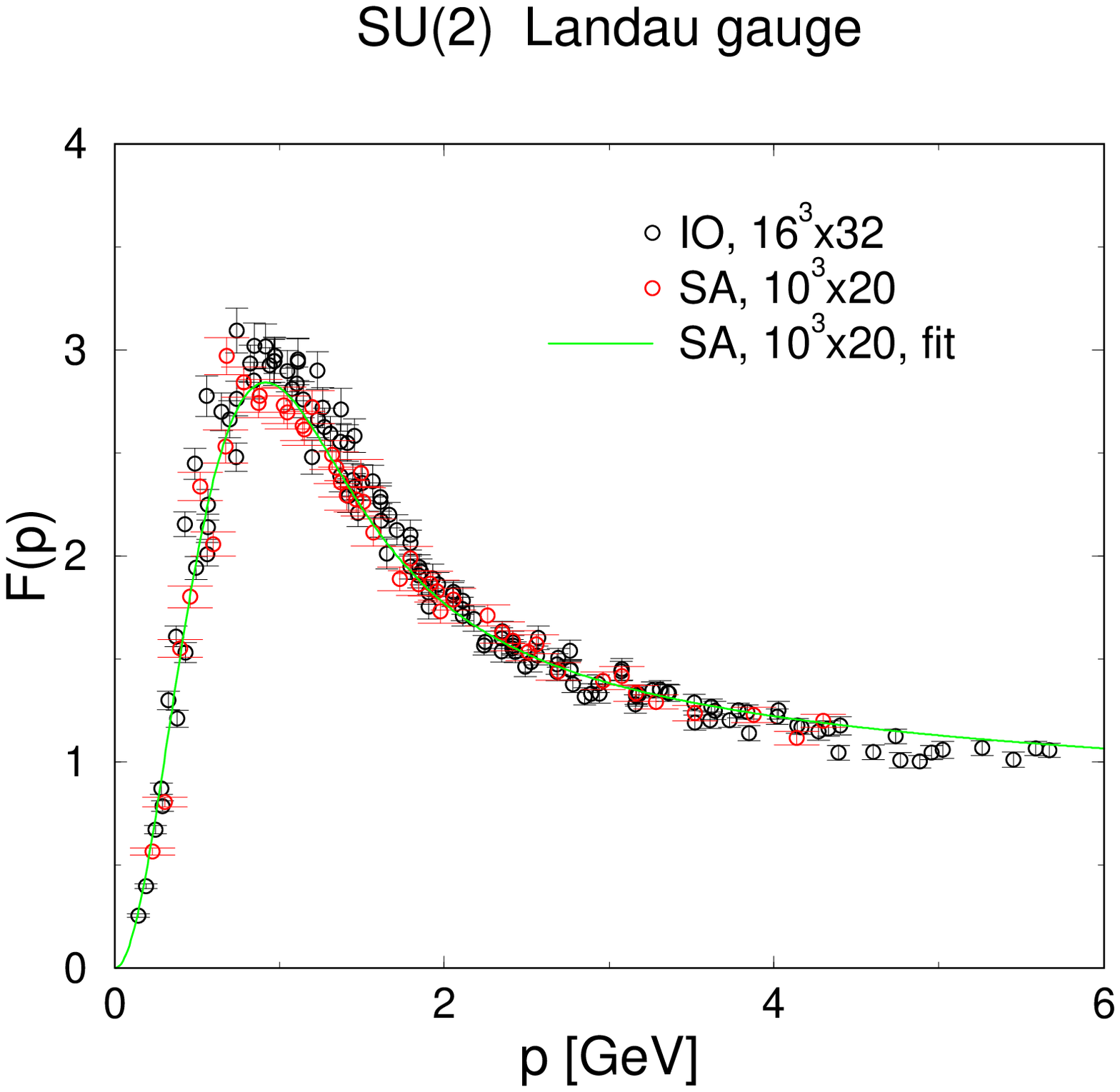}
}
\caption{The gluonic form factor $F(p^2)$ for a $10^3\times 20$ lattice 
   in the gauge IO and SA, respectively (left panel) and compared with 
   previous results (IO, $16^3\times 32$) (right panel). } 
\label{fig:20}
\end{figure}
Finally, let us check how strongly the gluonic 
form factor $F(p^2)$ depends on the choice of gauge, i.e. on the sample 
of maxima of the variational condition (\ref{eq:9}) selected by the algorithm. 
For this purpose, we adopt an extreme point of view by comparing the gauge 
implemented by the iteration over-relaxation (IO) algorithm with the 
gauge obtained by simulated annealing (SA). The results obtained for 
the form factor 
in both cases are shown in figure~\ref{fig:20}. We find, in agreement 
with~\cite{cuc97}, that, in the case of the gluonic form factor, 
the Gribov noise is comparable with the statistical noise for data generated 
with $\beta \in [2.1,2.5]$ (scaling window).

\section{ Gluonic spectral functions } 

\subsection{ The spectral density } 

Let us consider the following spectral representation of the trace of the 
Euclidean propagator $D(p^2)$ 
\be 
D(p^2) \; = \; \int _0 ^\infty 
dm \, \frac{ \rho (m) }{ p^2 \, + \, m^2 } \; , 
\label{eq:p20} 
\en 
where $\rho (m) $ is the spectral density. For example, the 
propagator of free particle with mass $m_p$ is represented 
by a spectral density $\rho (m) = \delta (m - m_p)$. 
In general, the spectral density carries information on the strength 
with which single particle states contribute to the correlation 
function of interest. Hence, the spectral density must be positive 
if the space of physical particles is considered. Note, however, that 
this constraint must be abandoned if un-physical (while gauge dependent) 
correlation functions, e.g.~the gluon propagator, is investigated. 
Moreover, the so-called negative norm states play an important role 
in Yang-Mills theory to circumvent the cluster decomposition theorem, 
what is inevitable to accomplish confinement of the theory. 
For a more detailed discussion of these issues see~\cite{alk00}. 

\vskip 0.3cm 
In order to study the contribution of hypothetical negative norm states 
to the gluon propagator, we will calculate the spectral density from 
the gluonic form factor
\be 
F(p^2) \; = \; \int dm \, \rho (m) \, \frac{ p^2 }{ p^2 \, + \, m^2 } \; . 
\label{eq:p21} 
\en 
Taking the derivative of $F(p^2)$ (\ref{eq:21}) with respect to $p$, 
one finds 
\be 
F^\prime(p^2) \; = \; 2p \int dm \, \rho (m) \frac{ m^2 }{ \bigl( 
p^2 \, + \, m^2 \bigr)^2 } \; . 
\label{eq:p22} 
\en 
Hence, one immediately concludes that the form factor would be a monotonic 
function of the momentum if only positive norm states contribute, 
i.e.~if $\rho (m) \ge 0$. 
An inspection of the results of the previous section shows that 
this is not the case for the gluon propagator. 

\vskip 0.3cm 
Further insights are provided by the sum rules which are obtained 
from a large momentum expansion of the form factor (\ref{eq:p21}), i.e. 
\be 
F(p^2) = \sum _{n=0}^\infty c_n \frac{1}{p^{2n} } \; , \hbo 
c_n = (-1)^n \int dm  \; \rho (m) \; m^{2n} \; . 
\label{eq:p22b} 
\en 
where $n<n_c$ is restricted from above in order to guarantee the existence 
of the momentum integrals. In particular, we observe that 
\be 
\lim _{p^2 \rightarrow \infty } F(p^2) \; = \; \int dm \; \rho (m) \; . 
\label{eq:p22c} 
\en 
There is is a substantial difference between the spectral function of 
a free massive particle and the one of Yang-Mills theory: while in 
the free particle case the form factor approaches unity at large 
momentum transfer, we know from perturbation theory that the 
Yang-Mills form factor vanishes at large momentum transfer. In view of 
(\ref{eq:p22c}), the crucial observation is that the free particle 
form factor $F(p^2 \rightarrow \infty )=1$ is consistent with a 
positive definite spectral function, while the vanishing form factor 
$F(p^2 \rightarrow \infty )=0$ in the Yang-Mills case implies that 
unless $\rho (m)$ vanishes identically it must change sign. 

\vskip 0.3cm
The remaining two subsections are devoted to numerically estimate 
the spectral function $\rho (m)$ from the form factor data in oder 
to get insights into the qualitative behavior of the spectral function 
of a confining theory.

\subsection{ The generalized Maximal Entropy Method } 
\label{sec:mem} 

In order to extract the spectral function $\rho (m)$ from our gluon 
propagator obtained in the lattice simulations we employ 
the Maximal Entropy Method (MEM). There is a wide span of applications for 
the MEM ranging from image modeling to solid state physics~\cite{jar96}. 
It was recently applied to reconstruct the spectral density of 
mesons from correlation functions obtained in lattice 
calculations~\cite{wet00}. Since these applications use the positivity 
of the spectral density, a slight generalization of the MEM is 
necessary for our purposes. 

\vskip 0.3cm 
Given the definition 
\be 
f_{\mathrm MEM} (p^2) \; := \; \int _0 ^\infty dm \, \rho (m) \frac{ p^2 }
{ p^2 \, + \, m^2 } \; , 
\label{eq:p23} 
\en 
we define the MEM potential functional by 
\be 
V [\rho ] \; = \; \int dp \; \frac{1}{ \sigma (p) } \biggl[ 
F(p^2) \, - \, f_{\mathrm MEM} (p^2) \biggr] ^2 \; , 
\label{eq:p24} 
\en
where $\sigma (p) $ denotes the standard deviations of the measured 
values $F(p^2)$. For given $F(p^2)$, minimization of the functional 
$V [\rho ] $ with respect to the spectral function $\rho (m)$ 
defined by (\ref{eq:p23}) corresponds to a 
least square fit and would result in the optimal choice for 
$\rho (m)$ representing the data. However, it appears that 
substantial changes of the function $\rho (m)$ produce 
minor changes of the function $f_{\mathrm MEM} (p^2) $ (\ref{eq:p23}) 
which are comparable in size with the error bars of the measured 
function $F(p^2)$. In practice, the numerical algorithm which tries 
to minimize the potential (\ref{eq:p24}) is unstable. 

\vskip 0.3cm 
For circumventing this problem, one defines a MEM action functional 
\be 
S [\rho ] \; = \; \alpha \, S_{\mathrm entropy }[\rho ]  \; 
+ \; V [\rho ] 
\label{eq:p25} 
\en 
where the parameter $\alpha $ regulates the influence of the 
entropy function $S_{\mathrm entropy }[\rho ]$ (to be specified below) 
on the minimum of the functional $S[\rho ]$. 
The action functional (\ref{eq:p25}) is minimized instead of the potential 
(\ref{eq:p24}). 
Thereby, the entropy factor $S_{\mathrm entropy }[\rho ] $ 
is a measure for the deviation of the spectral density $\rho (m)$ 
from a default density $\rho _{\mathrm def} (m)$. 
Whenever the potential term (\ref{eq:p24}) is not conclusive on the 
precise form of the spectral density due to the size of the error bars, 
the numerical algorithm minimizes the deviations of $\rho (m)$ 
from the default model specified by $\rho _{\mathrm def} (m)$ . 
The entropy factor $S_{\mathrm entropy }[\rho ] $ 
is also used to encode constraints on the function $\rho (m)$. 
For instance, if one would like to insist on the positivity of the 
spectral function, the generic choice is~\cite{jar96} 
\be 
S_{\mathrm entropy }[\rho ] \; = \; \int dm \; \biggl[ 
\rho (m) \, \log \frac{ \rho (m) }{ \rho _{\mathrm def}(m) } \; - \; 
\rho (m) \, \biggr] \; , 
\label{eq:p26} 
\en 
since the functional (\ref{eq:p26}) has a unique minimum for 
$\rho (m) = \rho _{\mathrm def}(m)$. 

\vskip 0.3cm 
For each value $\alpha $ a minimum of the action (\ref{eq:p25}) 
exists for properly chosen entropy functionals. 
Let $\rho _\alpha (m)$ denote the function which minimizes 
(\ref{eq:p25}), and let $s(\alpha )$ be the corresponding minimum value. 
The {\it most probable } spectral function is then defined by 
\be 
\rho _{av} (m) \; = \; \frac{1}{\cal N} \int d \alpha \; \rho _\alpha (m) 
\; \exp \biggl\{ - s(\alpha ) \biggr\} \; , \hbo 
{\cal N} : = \int d \alpha 
\; \exp \biggl\{ - s(\alpha ) \biggr\} \; . 
\label{eq:p27} 
\en 
Of course, it is mandatory to check that $f_{\mathrm MEM} (p^2)$ 
(\ref{eq:p23}) which is calculated with $\rho _{av} (m) $ does 
indeed represent the data $F(p^2)$ within statistical error bars. 

\begin{figure}[h]
\centerline{
\epsfxsize=0.5\linewidth 
\epsfbox{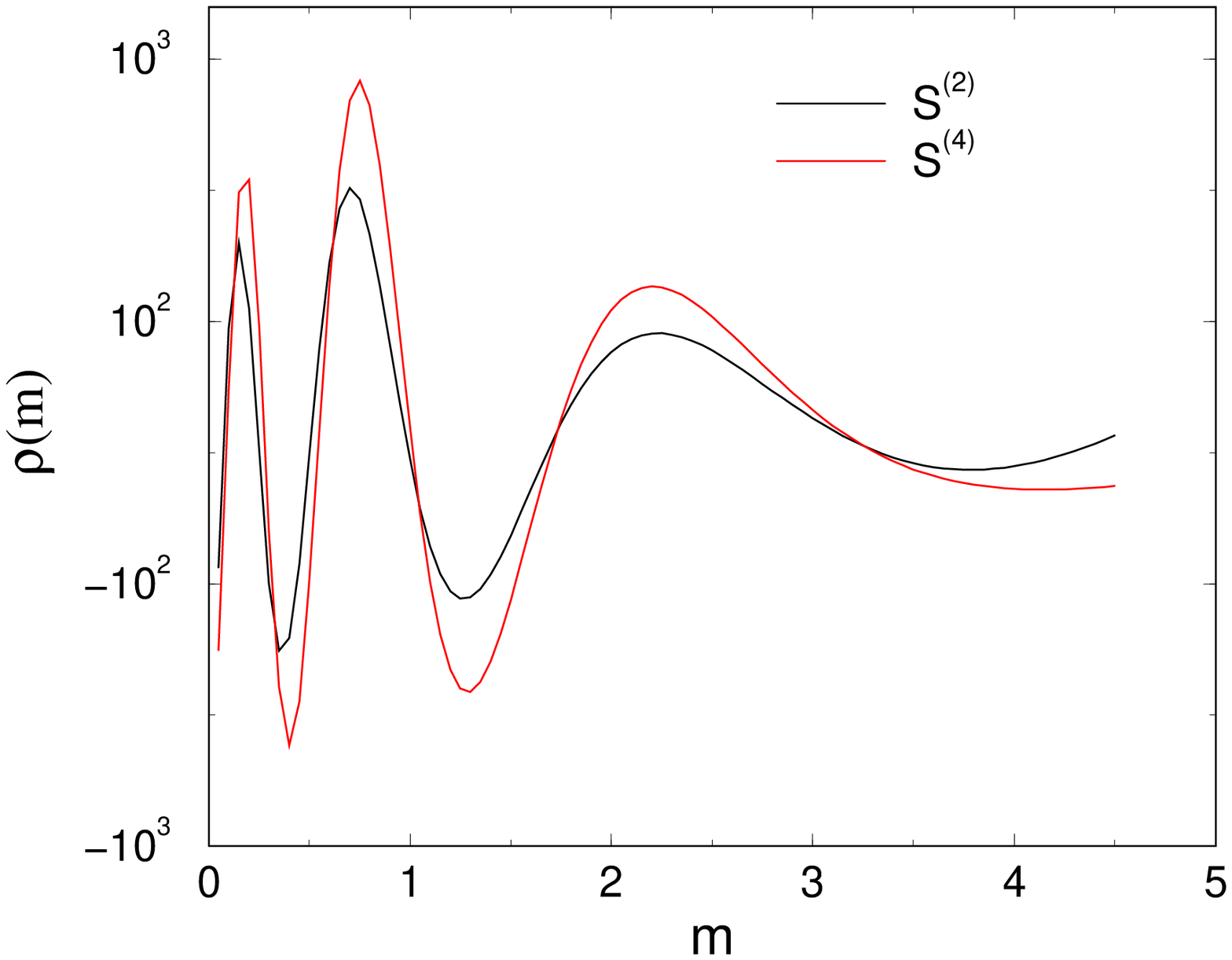}
\epsfxsize=0.46\linewidth 
\epsfbox{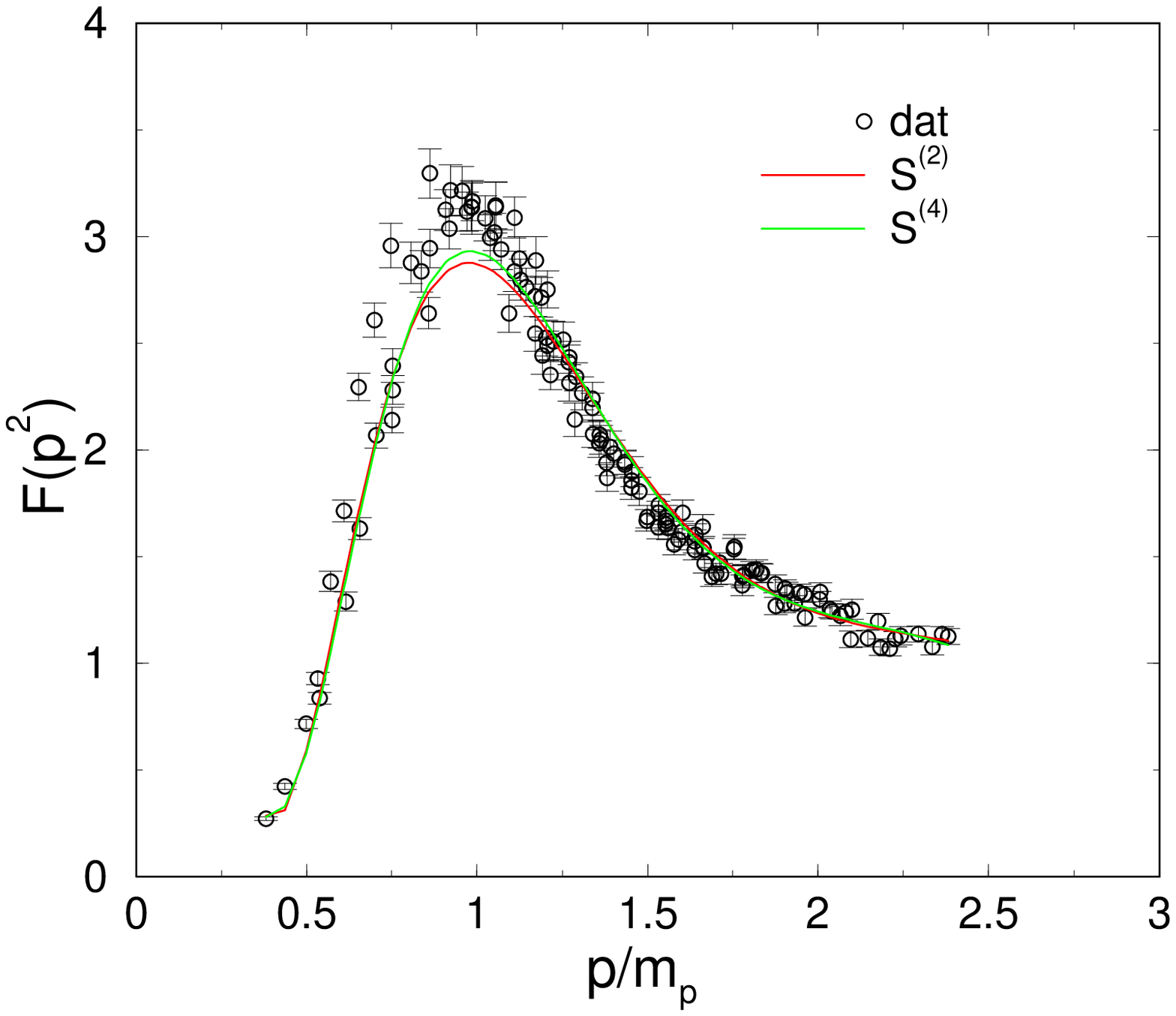}
}
\caption{The spectral functions obtained with the entropy 
   functionals $S^{(2)}$ and $S^{(4)}$, respectively, (left panel) and the 
   corresponding form factors (right panel). } 
\label{fig:30} 
\end{figure}
\vskip 0.3cm 
Since we must abandon the positivity constraint, we cannot use the 
standard entropy functional (\ref{eq:p26}). We will here explore 
the two functionals   
\bea 
S^{(2)}_{\mathrm entropy }[\rho ]  &=& \int dm \; \biggl[ 
\frac{ d \rho (m) }{ dm } \biggr]^2 \; , 
\label{eq:p28} \\ 
S^{(4)}_{\mathrm entropy }[\rho ]  &=& \int dm \; \biggl[ 
\rho (m) \, - \,  \rho _{\mathrm def}(m) \, \biggr]^2 \; , 
\label{eq:p29} 
\ena 
where we use a ''default model'' which suppresses the density $\rho (m)$ 
at small and at high values of $m$, e.g. 
\be 
\rho _{\mathrm def}(m) \; = \; \frac{ m^2 }{ m^4 + (2 \, m_p)^4 } \; , 
\label{eq:p30} 
\en 
\begin{figure}[h]
\centerline{
\epsfxsize=0.5\linewidth 
\epsfbox{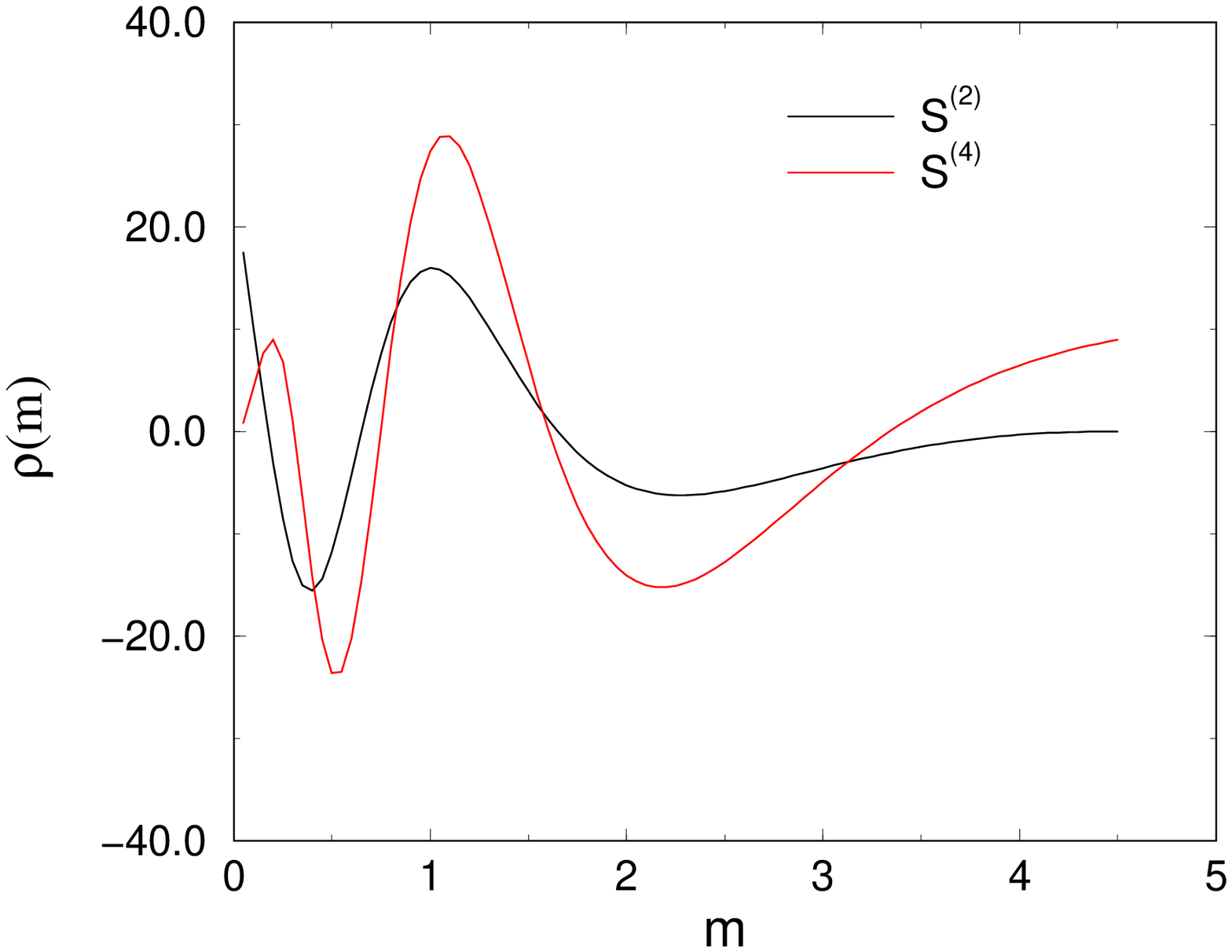}
\epsfxsize=0.48\linewidth 
\epsfbox{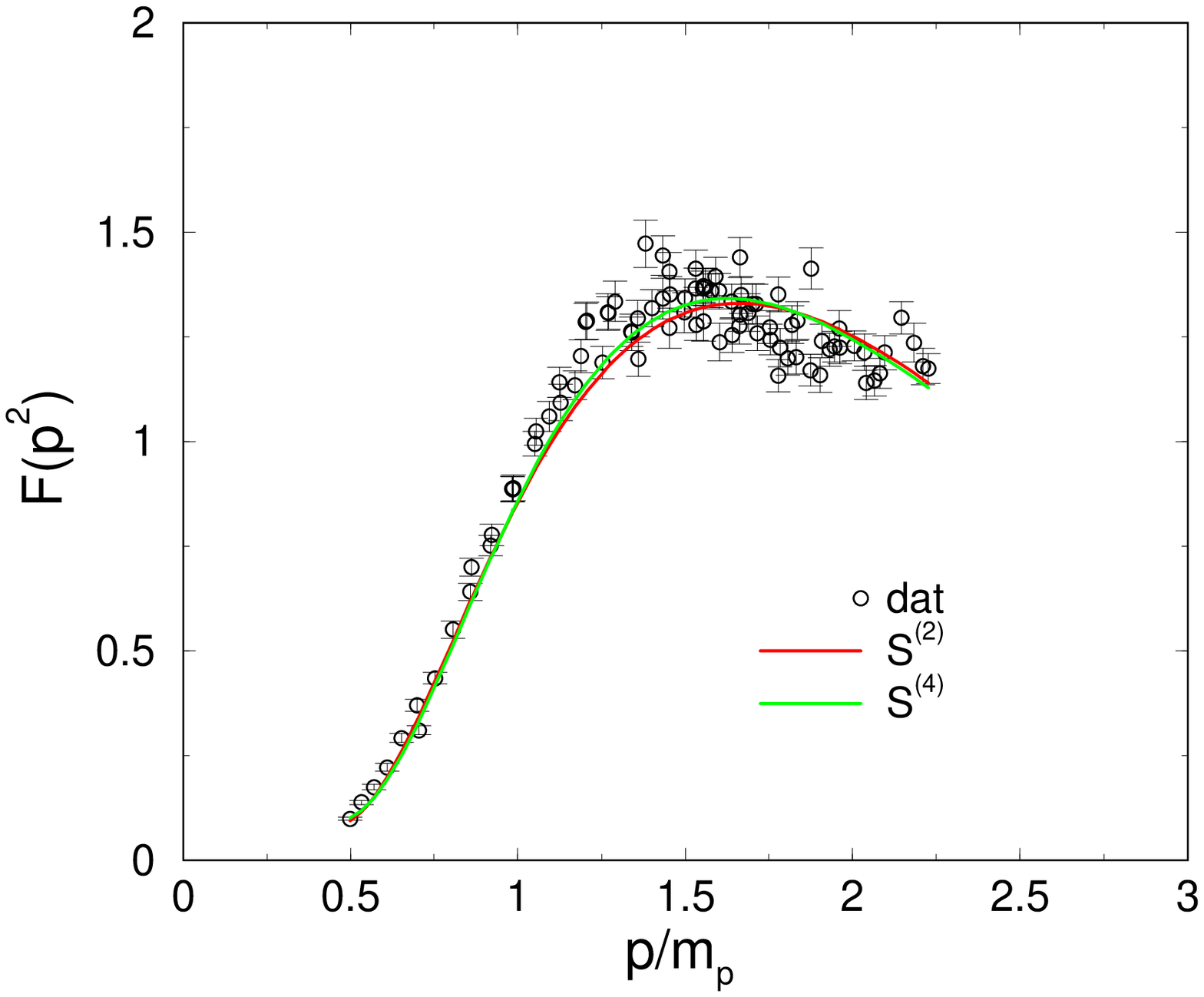}
}
\caption{ Same as figure \ref{fig:30} for the case of the non-confining 
   Yang-Mills ensemble. } 
\label{fig:31} 
\end{figure}
where the mass scale is chosen to be $m_p = 1 \, $GeV. 
The first functional suppresses large gradients and, hence, ensures 
that the spectral function is smooth. The second functional allows 
for large gradients, but minimizes the deviation of the spectral function 
from that of a ''default model''. 
The two functionals have been tested for the case of the form factor 
of a free particle (see appendix \ref{app:b}). The spectral function of the 
cases (\ref{eq:p29}) and (\ref{eq:p30}), respectively, have been 
compared with the result produced with the standard entropy functional 
(\ref{eq:p26}). Since less information is supplied for the choices 
(\ref{eq:p29}, \ref{eq:p30}), the exact infinite volume density is 
reproduced to less accuracy than in the case using the 
standard entropy functional (\ref{eq:p26}). These results stress the 
importance of the physical constraints on the spectral function. 
We point out that, unless additional information on the spectral function 
of the gluon form factor is obtained and is used to constrain the 
latter, one must 
accept the uncertainty which becomes visible if two different entropy 
functionals (such as (\ref{eq:p29}) and (\ref{eq:p30}) are used. 

\subsection{ MEM fit of the lattice form factor } 
\label{sec:lan} 

In a first step, we do not attempt to supplement additional physical 
information to the MEM approach, but produce a direct MEM fit of 
the form factor data by using the two completely different 
entropy functionals (\ref{eq:p28}), (\ref{eq:p29}) in order to get a 
clue about the residual freedom in the choice of the spectral function 
$\rho (m)$. Our findings for the spectral function corresponding to the 
form factor $F(q^2)$ of the gluon propagator in Landau gauge is 
summarized in figure \ref{fig:30}. The MEM approaches which employ the 
entropy functionals (\ref{eq:p28}), (\ref{eq:p29}) produce qualitatively the 
same result for the spectral function: $\rho (m) $ is a rapidly 
oscillating function of $m$. 

\begin{figure}[h]
\centerline{
\epsfxsize=0.5\linewidth 
\epsfbox{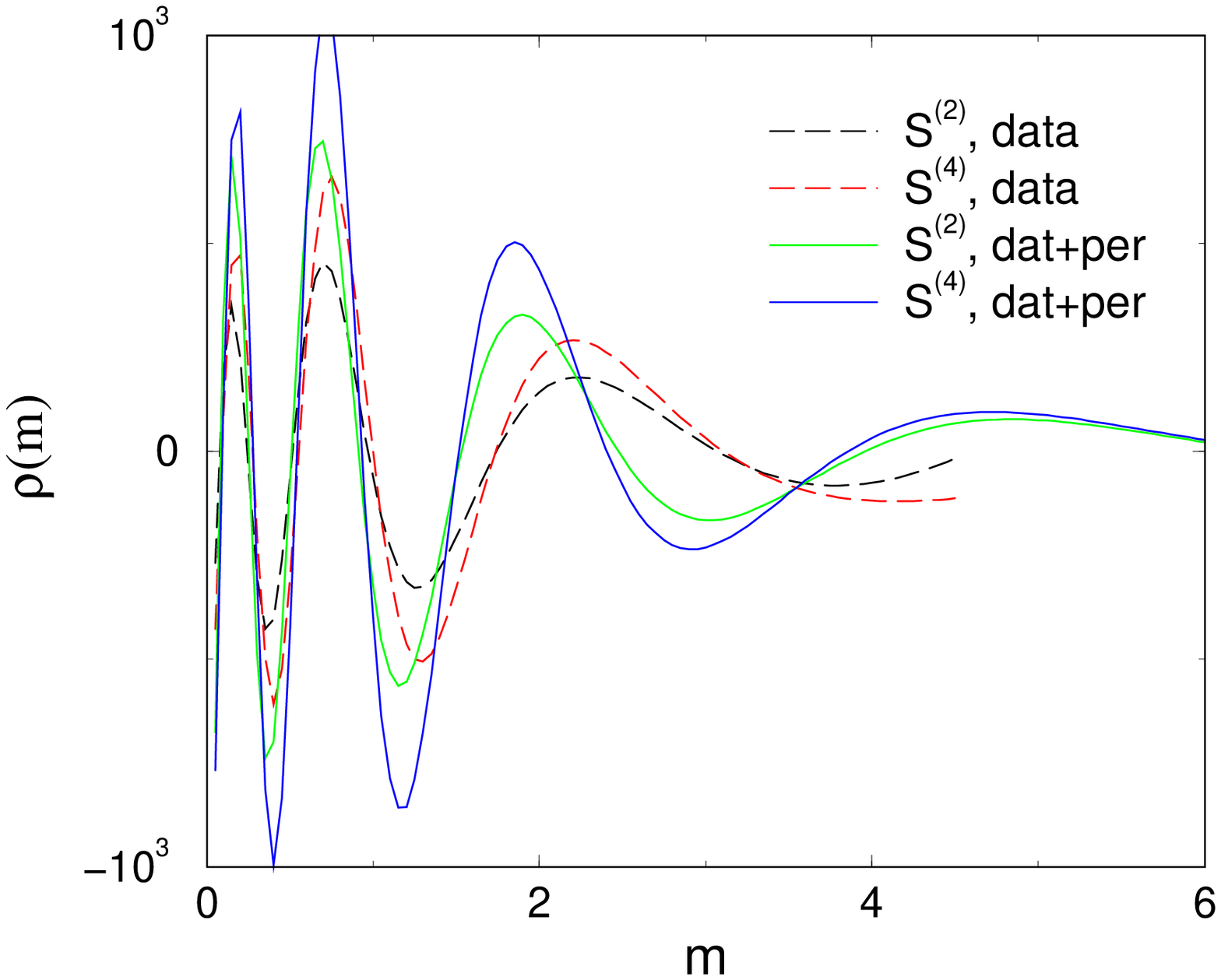}
\epsfxsize=0.48\linewidth 
\epsfbox{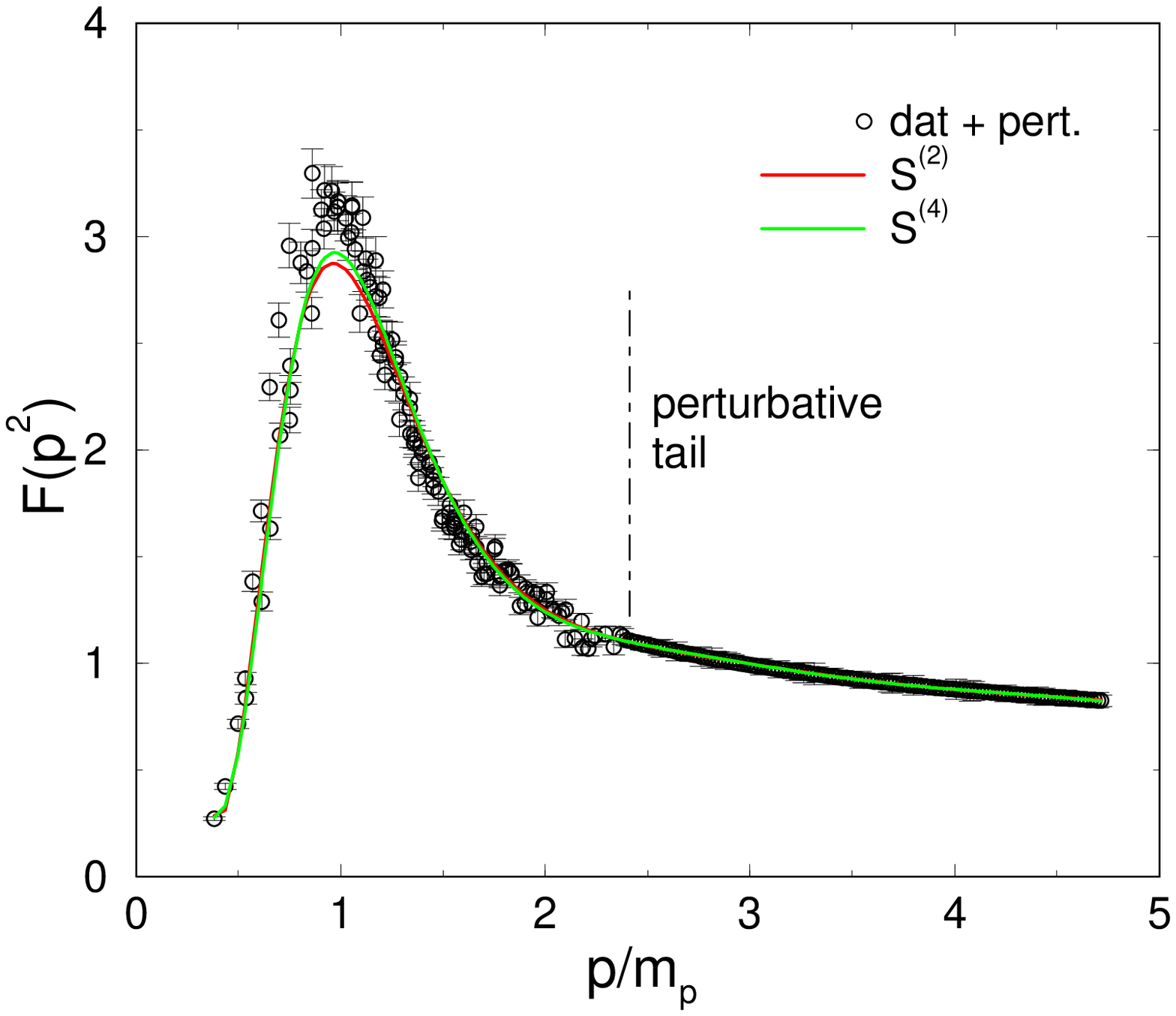}
}
\caption{ Same as figure \ref{fig:30} for the case that the perturbative 
   tail has been added (full Yang-Mills ensemble). } 
\label{fig:32} 
\end{figure}
\vskip 0.3cm 
We compare these findings with the ones obtained from the modified 
Yang-Mills ensemble from which the center vortices have been removed 
by the method described in the previous section. As shown there, the 
modified ensemble is non-confining. 
Since in this case, as in the case of the full ensemble, 
the form factor is non-monotonic due to the perturbative tail, negative 
norm states are inevitable. The MEM fit to the data indeed 
shows an oscillation of the spectral function $\rho (m)$ the 
amplitude of which is, however, orders of magnitude smaller than in the 
case of the full Yang-Mills theory (see figure~\ref{fig:31}). 

\vskip 0.3cm
Finally, we will assume that the perturbative regime is already 
approached at the upper limit of the momentum range which has been 
explored by the lattice calculation so far. Therefore, we will 
artificially extend the momentum range of the form factor by 
attaching the perturbative tail to the high momentum form factor 
obtained from the lattice data. In addition, we add a Gaussian noise 
to the perturbative tail. The noise is of the same order of magnitude 
as the statistical error bars, and specifies the degree of freedom which 
the MEM method might exploit to reproduce the form factor data. 

\begin{figure}[h]
\centerline{
\epsfxsize=0.5\linewidth 
\epsfbox{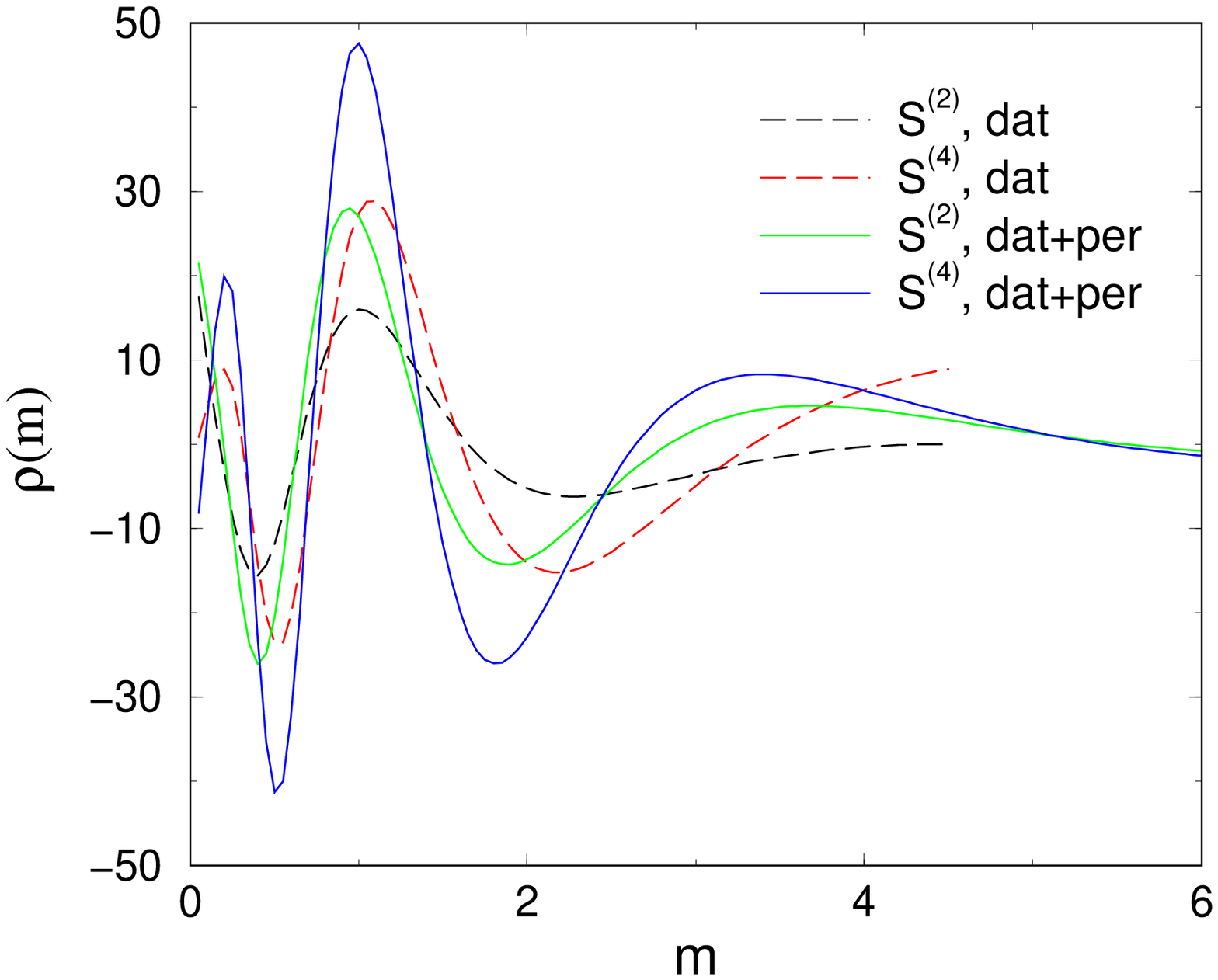}
\epsfxsize=0.48\linewidth 
\epsfbox{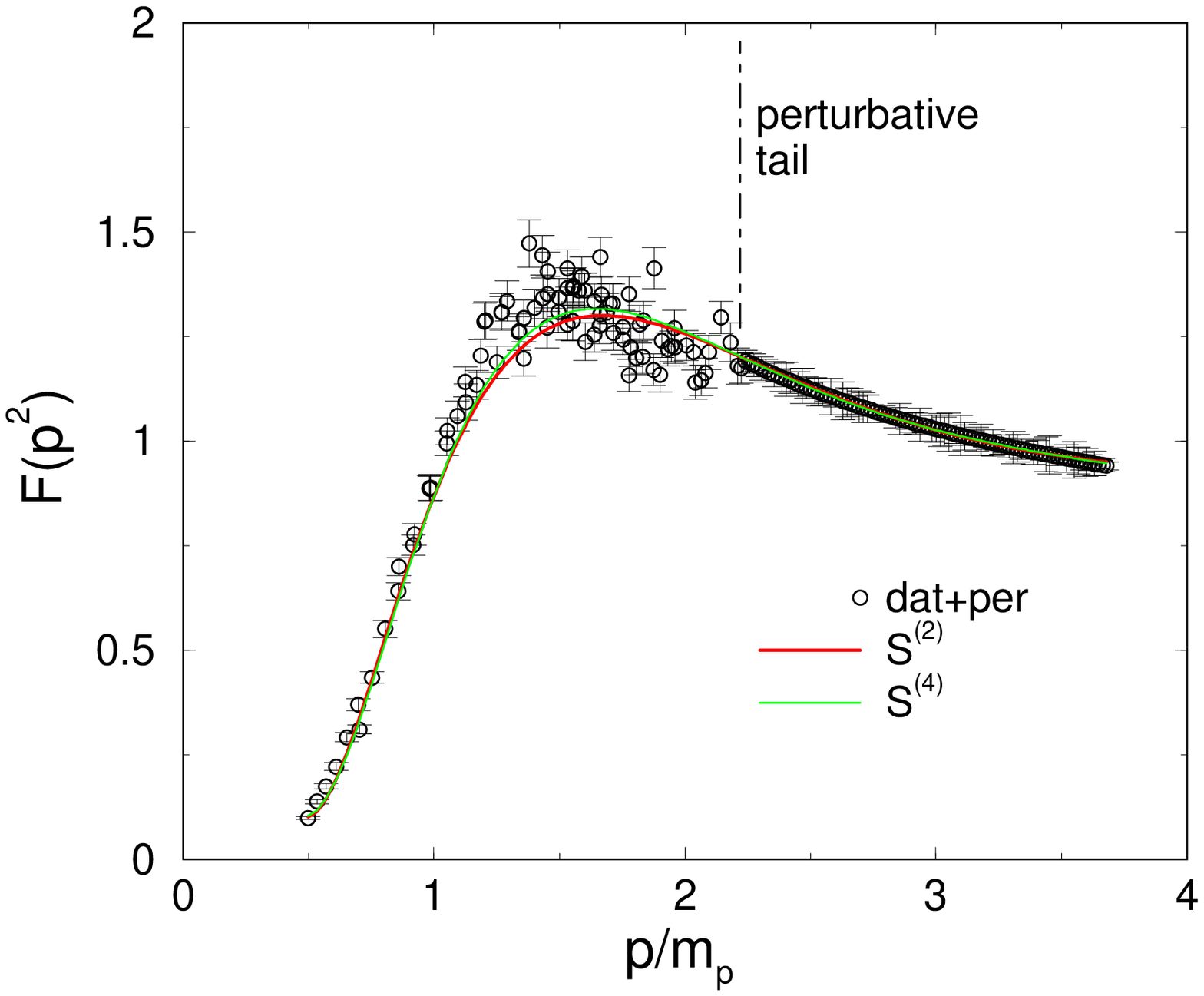}
}
\caption{ Same as figure \ref{fig:30} for the case that the perturbative 
   tail has been supplemented to the lattice data (modified non-confining 
   ensemble).} 
\label{fig:33} 
\end{figure}
\vskip 0.3cm 
Figure \ref{fig:32} shows the result for the spectral function for the 
case of the full Yang-Mills gluonic form factor. Most important is 
the observations that the oscillations are stable in the position. There 
is a enhancement of the amplitudes which is naturally expected if 
the momentum range of the data grows. 

\vskip 0.3cm 
We also repeated the analysis of the spectral function for the case 
of the non-confining ensemble (see figure \ref{fig:33}). 
Also in this case, we find that the 
attachment of the known perturbative behavior at large momentum transfer 
does not qualitatively change the behavior of the spectral function 
$\rho (m)$, but results in an enhancement of the amplitude. 

\subsection{ Infra-Red enhancement and negative norm states } 
\label{sec:ir} 

\begin{figure}[h]
\centerline{
\epsfxsize=0.5\linewidth 
\epsfbox{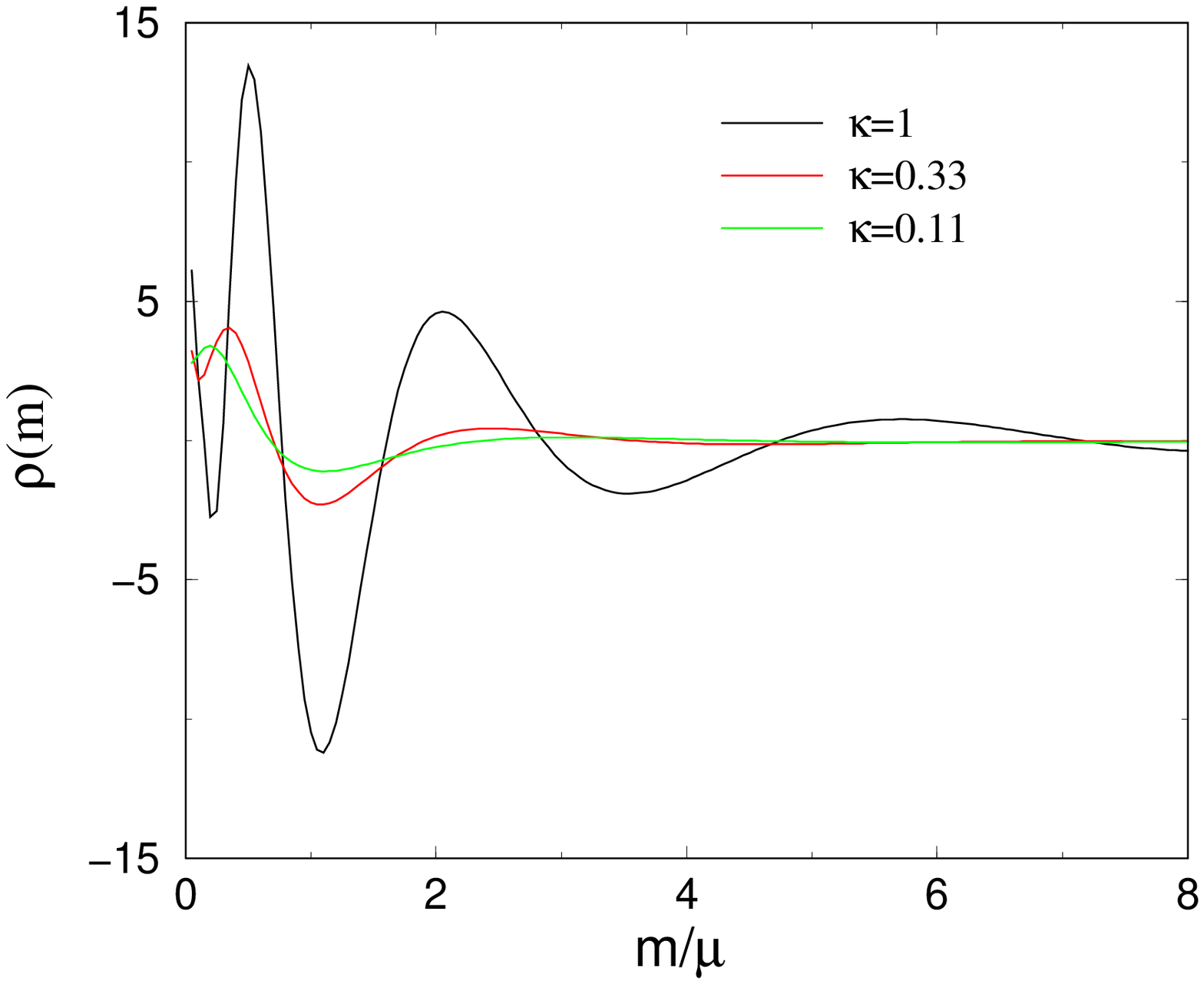}
\epsfxsize=0.48\linewidth 
\epsfbox{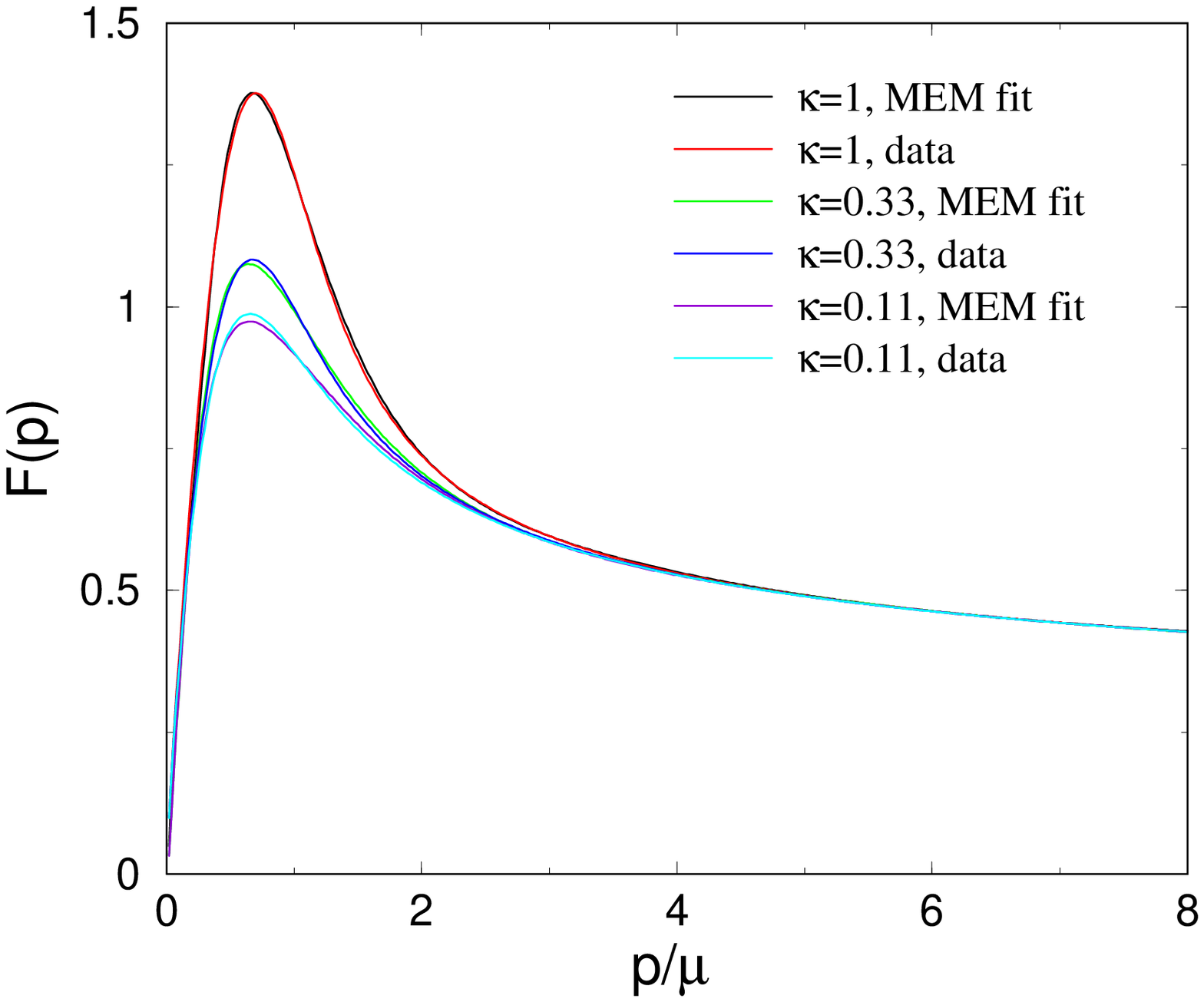}
}
\caption{ Spectral functions corresponding to the model form factors 
   $F_\kappa (p^2)$ (\ref{eq:p40}) for several values $\kappa $ 
   of the infra-red strength.  } 
\label{fig:40} 
\end{figure}
In order to detect the origin of the rapid oscillations in the 
spectral density $\rho (m)$ for the case of the full gluonic 
from factor (see figure \ref{fig:30}), we are guided by the mass fit 
(\ref{eq:26}) of the lattice study and take the class of functions, i.e. 
\be 
F_\kappa 
(p^2) \; = \; \frac{ p^2}{p^2+m_1^2} \biggl[ \frac{ \kappa 
\mu ^4 }{p^4+m_2^4} 
\; + \; \frac{1}{\left[ log \left(\frac{m_L^2}{\mu ^2 } + \frac{p^2}{\mu ^2 } 
\right) \right]^{13/22} } \biggr] \; , 
\label{eq:p40} 
\en 
where $m_1=0.64 \, \mu$, $m_2=m_L=\mu$ (varying $\kappa $), 
as a model for the gluonic form factor. Here, $\kappa $ controls 
the medium infra-red strength of the form factor. 
The MEM of the form factor fits which employ the entropy functional $S^{(2)}$ 
are shown in figure \ref{fig:40} for $\kappa = 1, 0.33, 0.11 $. 
These fits of the form factor fall on top 
of model functions $F_\kappa (p^2) $ within plotting accuracy. 
The corresponding spectral functions are also shown in \ref{fig:40}. 
One observes that amplitude of the oscillations of $\rho (m)$ 
increases if the infra-red strength, i.e. $\kappa $, rises.

\section{ Conclusions } 

We have studied the gluonic form factor $F(p^2)$ in Landau 
gauge and maximal center gauge, respectively, by means of the continuum 
extrapolated SU(2) lattice gauge theory. 
The gauge field has been defined from the adjoint link variables. 
It is well known~\cite{zwa94} that the Landau gauge condition (\ref{eq:9}) 
does not completely fix the gauge, but one must select a subset of 
configurations lying within the first Gribov horizon. We have studied 
two such possibilities: firstly, an iteration over-relaxation algorithm 
(IO gauge) which averages over the configurations within the first 
Gribov horizon, and, secondly, simulated annealing which selects the 
configuration belonging to the fundamental modular region (SA gauge). Within 
the statistical error bars, both methods have yielded the same gluonic form 
factor. Our results are also in agreement with the findings using the 
Laplacian Landau gauge~\cite{alex00} where the problem of Gribov 
ambiguities is evaded in an elegant way.

\vskip 0.3cm 
Close to zero momentum transfer, the gluonic form factor is mass 
dominated where the mass is $\approx 650 \pm 20 \,$MeV
(we used $\sqrt{\sigma } = 440 \,$MeV as reference). The 
uncertainty is due to statistical errors as well as due to 
Gribov ambiguities, i.e. using IO and SA-gauge, respectively. 
Our numerical results are stable against a variation of the 
UV-cutoff (i.e.$\pi/a$) and the physical volume by roughly an order 
of magnitude (see table \ref{tab:1}). The data points of the form 
factor in the far infra-red were obtained by using lattice sizes 
of $L \approx 9 \, $fm. This volume corresponds to a topological mass 
$2\pi / L \approx 140 \, $MeV which is much smaller than the dynamical 
gluon mass $m_1 \approx 650 \,$MeV. This explains the small finite 
size effects on the form factor observed by our numerical simulation.

\vskip 0.3cm 
In the medium momentum range we have found a rather pronounced peak while 
at high momenta our numerical data nicely reproduce the result 
from perturbative Yang-Mills theory. Our numerical data are well 
fitted over the whole momentum range by the formula (\ref{eq:26}) 
which might be useful for further phenomenological oriented 
investigations. 

\vskip 0.3cm 
Our SU(2) gluonic form factor in Landau gauge shows the same 
qualitative behavior than its SU(3) counterpart using Landau 
gauge~\cite{bon01} or using Laplacian Landau gauge~\cite{alex00,alex01}.

\vskip 0.3cm 
A focal point of our studies is the information on quark confinement 
which might be encoded in the gluon propagator in Landau gauge. 
By removing the confining vortices from the ensemble by hand, we are 
left with an ensemble which does not confine quarks (see figure 
\ref{fig:2} left panel). After the implementation of Landau gauge, 
we have seen that a good deal of strength is removed in the medium 
momentum range. We have therefore established a relation of the infra-red 
strength of the gluonic form factor in Landau gauge and quark confinement. 

\vskip 0.3cm 
We have compared our result for the gluonic form factor with that obtained 
by solving a truncated set of Dyson-Schwinger equations  (DSE) in continuum 
formulated Yang-Mills theory~\cite{cfi01}. The DSE result 
is in qualitative agreement with our findings, but does not reproduce 
the data on a quantitative level (except in high momentum, perturbative 
regime): the power law behavior of the form factor in the vicinity 
of zero momentum is different, and the peak of the intermediate momentum range 
is less pronounced given that both approaches give same results in 
perturbative momentum regime. 

\vskip 0.3cm 
Finally, we have studied the gluonic spectral functions which reproduce 
the numerical data for the gluon form factor. Using 
sum rule techniques, it becomes clear that the spectral function 
necessarily comprises negative parts. Using a generalized 
version of the Maximum Entropy Method (MEM), we have found that this is 
the case for the full form factor as well as for the form factor 
obtained for the non-confining ensemble. At a quantitative level, 
the spectral function corresponding to the full form factor 
shows large amplitude fluctuations while the amplitude of the 
oscillations in the case of the spectral function obtained from the 
non-confining ensemble are moderate.

\vspace{1cm}
{\bf Acknowledgments.} Helpful discussions with R.~Alkofer, J.~C.~R.~Bloch, 
C.~Fischer, L.~v.~Smekal and P.~Watson are greatly acknowledged.

\appendix
\section{ Gauge fixing on the lattice } 
\label{app:a}

In practical calculations, the following convenient procedure, which 
circumvents the explicit evaluation of the Faddeev Popov determinant, 
is adopted: 
a number of $\{U_\mu (x)\}_{i = 1 \ldots n}$ of statistically 
independent ensembles are generated by standard techniques. Each of these 
ensembles is then subject of the implementation of the gauge 
condition (\ref{eq:9}), i.e. 
\be 
\{U _\mu (x) \} _i \rightarrow \{U^\Omega _\mu (x)\}_i \; , \; \forall i \; .
\label{eq:e1} 
\en 
Depending on the algorithm, a single candidate of all possibile maxima 
of the variational gauge condition is randomly selected 
(see discussion in subsection~\ref{sec:gf}). 
Independently of the procedure which selects the gauge transformation 
$\Omega (x)$, a 
particular member  $\{U^\Omega _\mu (x)\}_k, \, k\in\{1\ldots n\}$ 
of the gauge fixed configurations 
is generated with a frequency proportional to the probability distribution 
of the gauge fixed sub-manifold. An estimator of a quantity $A$ 
is obtained by 
\be 
\biggl\langle A^\Omega \biggr\rangle \approx \frac{1}{n} \sum _{i=1}^n 
\biggl[A\biggr]_{\{U^\Omega_\mu (x)\}_i} \; . 
\label{eq:e2} 
\en 
Let us consider the particular example that $A$ is a gauge invariant 
combination of fields, $A=A^\Omega$. One 
obtains with (\ref{eq:e2}) 
\be 
\biggl\langle A \biggr\rangle \approx \frac{1}{n} \sum _i^n 
\biggl[A\biggr]_{\{U_\mu (x)\}_i} \; = \; 
\frac{1}{n} \sum _i^n 
\biggl[A ^\Omega \biggr]_{\{U_\mu (x)\}_i} \; = \; 
\frac{1}{n} \sum _i^n 
\biggl[A\biggr]_{\{U^\Omega_\mu (x)\}_i} \; . 
\label{eq:e3} 
\en 
Hence, gauge invariant 
quantities which are calculated from the gauged configurations evidently 
coincide with the ones obtained from un-fixed configurations.

\section{ Generalized MEM method (example) } 
\label{app:b}
\begin{figure}[t]
\centerline{
\epsfxsize=0.5\linewidth 
\epsfbox{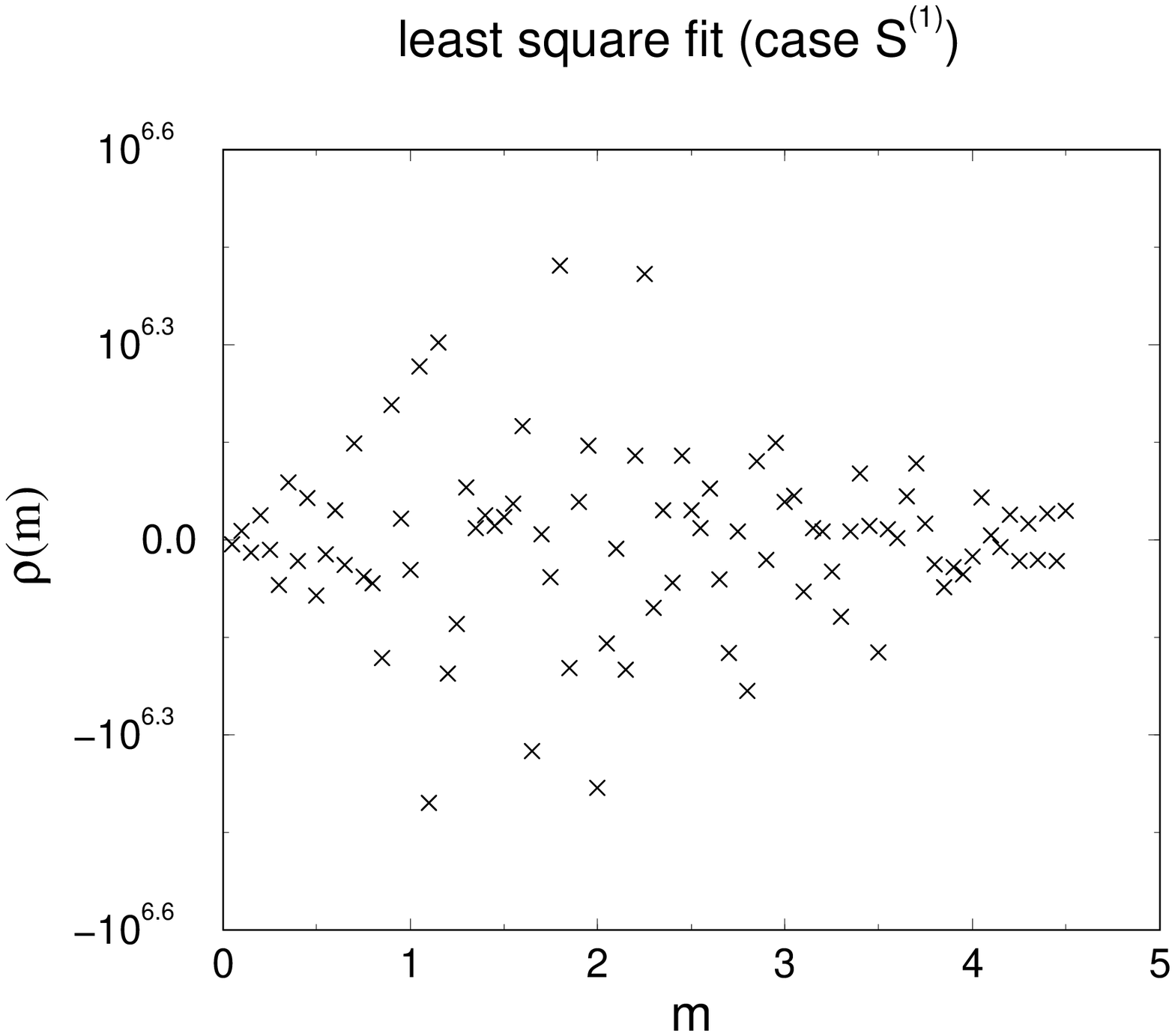}
\epsfxsize=0.5\linewidth 
\epsfbox{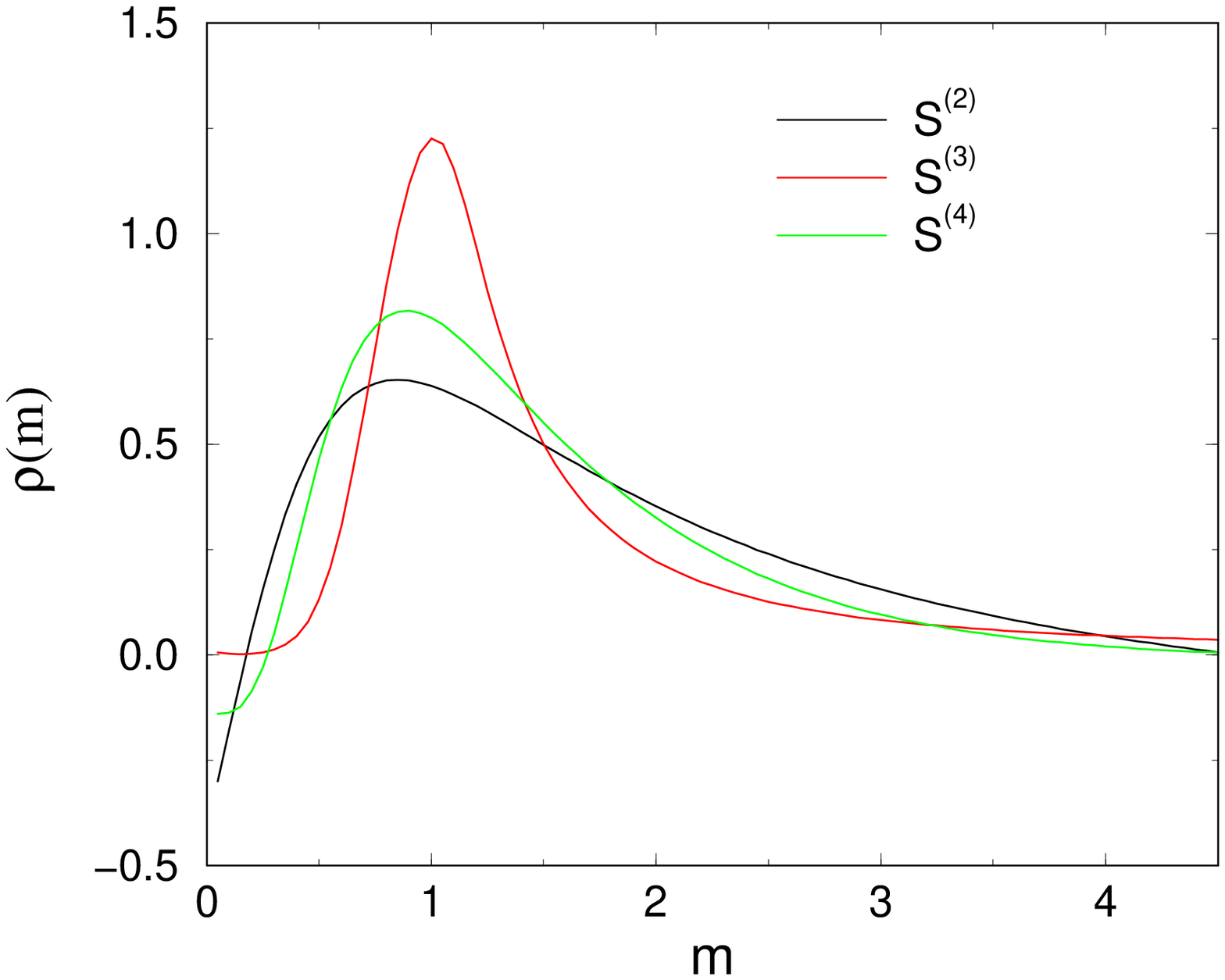}
}
\caption{ Possible spectral functions (differing by constraints) which 
   fit the form factor of a free massive particle. } 
\label{fig:a1}
\end{figure}
We will briefly investigate several proposals for the MEM entropy 
functional $S_{\mathrm entropy }[\rho ]  $ in (\ref{eq:p25}) for the 
case of the form factor of a free particle of mass $m_p$ where 
noise has been supplemented to the form factor by hand to simulate 
statistical uncertainties. The form factor is given by 
\be 
F(p^2) \; = \; \frac{ p^2 }{ p^2 + m_p^2} \; . 
\label{eq:pa1} 
\en 
In order to illustrate the sensitivity of the spectral function $\rho (m)$ 
to the constraints, which are incorporated in $S_{\mathrm entropy }[\rho ]  $, 
we will explore four different entropy functionals, i.e. 
\bea 
S^{(1)}_{\mathrm entropy }[\rho ]  &=& 0 \; , 
\label{eq:pa2} \\ 
S^{(2)}_{\mathrm entropy }[\rho ]  &=& \int dm \; \biggl[ 
\frac{ d \rho (m) }{ dm } \biggr]^2 \; , 
\label{eq:pa3} \\ 
S^{(3)}_{\mathrm entropy }[\rho ]  &=& \int dm \; \biggl[ 
\rho (m) \, \log \frac{ \rho (m) }{ \rho _{\mathrm def}(m) } \; - \; 
\rho (m) \, \biggr] \; , 
\label{eq:pa4} \\ 
S^{(4)}_{\mathrm entropy }[\rho ]  &=& \int dm \; \biggl[ 
\rho (m) \, - \,  \rho _{\mathrm def}(m) \, \biggr]^2 \; . 
\label{eq:pa5} 
\ena 
In order to perform the MEM program outlined in 
subsection~\ref{sec:mem}, all integrals are converted Riemann sums. 
We used 135 data points representing $F(p^2)$ and $N_m$ 
points to represent the spectral function $\rho (m)$.

\vskip 0.3cm 
The choice $S^{(1)}_{\mathrm entropy }$ corresponds to a least square 
fit of the spectral sum $f_{\mathrm MEM} (p^2)$ to the ''measured'' 
data. In this case, it turns out that the set of points $\rho (m)$ 
which represent the data set best is not a smooth function (see 
figure \ref{fig:a1} for the case $m_p=1$ and $N_m=90$).

\begin{figure}[t]
\centerline{
\epsfxsize=.8\linewidth 
\epsfbox{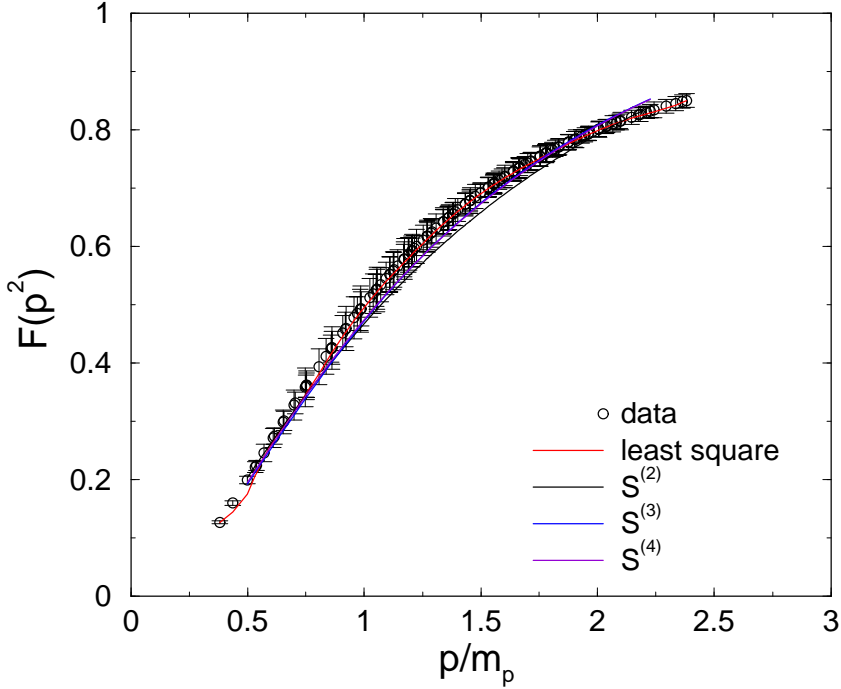}
}
\caption{ The form factor of a free massive particle fitted by 
   several spectral functions. } 
\label{fig:a2}
\end{figure}
\vskip 0.3cm 
The minimal requirement which we wish to incorporate in the spectral 
function $\rho (m) $ is that $\rho (m) $ is a smooth function of $m$. 
This is achieved be the entropy functional $S^{(1)}_{\mathrm entropy }$ 
(\ref{eq:pa3}) which disfavors functions with large gradients. The spectral 
function is also shown in figure \ref{fig:a1} where we have chosen 
$N_m=90$. One observes that a peak develops at $m=m_p$ approximating the 
exact infinite volume and zero noise spectral function $\rho _{exact} 
=  \delta (m-m_p) $. 

\vskip 0.3cm 
Let us compare, this result with the result of the standard approach 
demanding positivity. For this purpose, we use $S^{(3)}_{\mathrm entropy }$ 
(\ref{eq:pa4}). This choice for the entropy functional does not 
constrain gradients. The smoothness of $\rho (m)$ is here incorporated 
by minimizing the difference to a smooth default density 
$\rho _{\mathrm def}(m)$. Here, we used 
\be 
\rho _{\mathrm def}(m) \; = \; \frac{ m^2 }{ m^4 + (2 \, m_p)^4 } \; , 
\label{eq:pa6} 
\en 
which ensures that the spectral function vanishes for $m \rightarrow 0$ 
and  $m \rightarrow \infty $, respectively. This is the standard 
MEM approach to form factors with positive definite spectral functions. 

\vskip 0.3cm 
Finally, we use an ''Euclidean'' norm to measure the difference 
of the spectral function with the default model (see 
$S^{(4)}_{\mathrm entropy }[\rho ] $ (\ref{eq:pa5})). Thus, the 
constraint to positivity is abandon, and we, hence, expect that 
agreement of the MEM estimate of the spectral function with the exact 
result becomes worse. This is indeed observed (see figure \ref{fig:a1}). 

\vskip 0.3cm 
We point out that all above MEM suggestions for the spectral function
$\rho (m)$ fit the form factor $F(p^2)$ within the (artificial) statistical 
error bars. The best result for the spectral function is achieved 
with the entropy functional $S^{(3)}_{\mathrm entropy }[\rho ] $ 
(\ref{eq:pa4}) which incorporates the most severe constraints, i.e. 
smoothness and positivity.

\end{document}